\documentclass[aps,pra,reprint,amsmath,amssymb,twocolumn,superscriptaddress]{revtex4-2}
\usepackage[T1]{fontenc}
\usepackage[utf8]{inputenc}
\usepackage{graphicx}
\usepackage[normalem]{ulem}
\usepackage{dcolumn}
\usepackage{amsmath,amssymb,amsthm,mathtools}
\usepackage{bm}
\usepackage{soul}
\usepackage{xcolor}
\DeclareMathOperator{\tr}{tr}

\newcommand{\Ss}{\mathbb{S}}

\makeatletter
\newsavebox{\@brx}
\newcommand{\llangle}[1][]{\savebox{\@brx}{\(\m@th{#1\langle}\)}%
  \mathopen{\copy\@brx\mkern2mu\kern-0.9\wd\@brx\usebox{\@brx}}}
\newcommand{\rrangle}[1][]{\savebox{\@brx}{\(\m@th{#1\rangle}\)}%
  \mathclose{\copy\@brx\mkern2mu\kern-0.9\wd\@brx\usebox{\@brx}}}
\makeatother

\begin{document}

\title{Replica Theory of Spherical Boltzmann Machine Ensembles}

\author{Thomas Tulinski}
\affiliation{Laboratoire de Physique de l'\'Ecole Normale Sup\'erieure, PSL, CNRS UMR8023, Sorbonne Universit\'e, 24 rue Lhomond, 75005 Paris, France}
\author{Jorge Fernandez-de-Cossio-Diaz}
\affiliation{Institut de Physique Th\'eorique, Universit\'e Paris-Saclay, CNRS UMR3681, CEA, Gif-sur-Yvette, France}
\author{Simona Cocco}
\affiliation{Laboratoire de Physique de l'\'Ecole Normale Sup\'erieure, PSL, CNRS UMR8023, Sorbonne Universit\'e, 24 rue Lhomond, 75005 Paris, France}
\author{R\'emi Monasson}
\affiliation{Laboratoire de Physique de l'\'Ecole Normale Sup\'erieure, PSL, CNRS UMR8023, Sorbonne Universit\'e, 24 rue Lhomond, 75005 Paris, France}

\date{\today}

\begin{abstract}
Training in machine learning generally consists in finding one model, whose parameters minimize a data-dependent loss. Yet, empirical work shows that ensemble learning, an approach in which multiple models are sampled, can improve performance. Here, we provide an analytical framework to understand these observations in the case of Boltzmann machines, exploiting a duality between ensemble learning and large deviations of the free energy in spin-glass models. Replica calculations allow us to fully solve the case of spherical Boltzmann machine ensembles, and clarify when ensemble learning improves over standard loss minimization, in particular for nearly finite-dimensional data.   Our framework can also be applied to complex data distributions, in agreement with numerical simulations on deep networks.
\end{abstract}

\maketitle

\paragraph*{Introduction.}
Over the last decades, our understanding of disordered systems, in which some quenched variables $\boldsymbol J$ define the energy landscape $E$ of fast thermal variables $\boldsymbol{\sigma}$, has considerably improved \cite{Parisi2023, Charbonneau2023}. Such progress was made possible by the use of dedicated statistical physics approaches, in particular the replica method, which describes the properties of $n\to 0$ copies of the $\boldsymbol{\sigma}$'s in the same quenched landscape. In this context the key quantity of interest is the average replicated partition function,
\begin{equation}\label{eq:replicas}
\overline{Z(\boldsymbol{J})^n}=\int\mathrm{d}{\boldsymbol J} \rho({\boldsymbol J})\, Z({\boldsymbol J})^n, \quad Z({\boldsymbol J})=\sum_{\boldsymbol{\sigma}} e^{-E(\boldsymbol{\sigma};{\boldsymbol J})}
\end{equation}
where $Z(\boldsymbol{J})$ is the partition function for the sample $\boldsymbol J$, and $\overline{\,\cdot\,}$ denotes the average over the probability density $\rho$ of the quenched variables. By allowing the number $n$ of replicas to take finite values, large deviations of the free energy can be explored, above ($n<0$) \cite{Dotsenko1994, Parisi2010} or below ($n>0)$ \cite{Coolen1993, Parisi2008} its typical value ($n\to 0$).

Energy-based models are also encountered in the field of  machine learning. Boltzmann Machines (BMs) \cite{Hinton2025} and their extensions offer a simple, yet powerful framework to learn arbitrarily complex data distributions, and generate new data with similar features. In this context, the thermal variables are the ${\boldsymbol{J}}$'s, while the quenched variables are the dataset ${\cal D}=(\boldsymbol{\xi}^1,...,\boldsymbol{\xi}^K)$. Denoting by $P_G$ the prior over ${\boldsymbol J}$, the posterior distribution  is given by
\begin{equation}\label{eq:post}
P_T({\boldsymbol J}|{\cal D}) = \frac {1}{{Y} (\mathcal{D};T)}\;\left[ P_{G}({\boldsymbol J}) \prod_{k=1}^K \frac{e^{- E(\boldsymbol{\xi}^k;{\boldsymbol J})}}{Z({\boldsymbol J})}  \right]^\frac 1T
\end{equation}
where $T$ is the training temperature. In most applications, $T$ is sent to $0^{+}$, which corresponds to maximum-a-posteriori (MAP) inference of ${\boldsymbol J}$.
For finite $T$, the distribution $P_T({\boldsymbol J}|{\cal D})$ defines an ensemble of models ${\boldsymbol J}$ beyond ${\boldsymbol J}_\text{MAP}$. The case $T=1$ corresponds in particular to the standard Bayesian setting, in which the normalization $Y$ in \eqref{eq:post} is called marginal likelihood \cite{MacKay2003}. 
As in thermodynamics, knowledge of $Y$ and of its derivative with respect to the temperature $T$ gives, in principle,  access to the diversity (entropy) and the mean performance (average energy) of the models drawn from the ensemble. 

Though the computation of $Y$, which involves high-dimensional integration over all models $\boldsymbol{J}$ is a priori very difficult, we show here how it can be done using a special duality, allowing us to transfer the extensive knowledge about disordered systems accumulated so far to the analysis of energy-based model ensembles. Indeed, formally integrating \eqref{eq:post} over $\boldsymbol{J}$ and comparing with \eqref{eq:replicas}, we see that $Y$ coincides, up to an irrelevant data-dependent factor, with $\overline{Z(\boldsymbol{J})^n}$ when $\rho({\boldsymbol J})$ is chosen to be $\propto P_{G}({\boldsymbol J})^{1/T}\, \exp(-\sum_k E(\boldsymbol{\xi}^k;{\boldsymbol J})/T)$ and the number of replicas is $n=-K/T$.
Informally speaking, this duality turns the study of the ensemble of models $\boldsymbol{J}$ into the characterization of the data $\boldsymbol{\sigma}$ they generate. Besides its technical advantages, it is therefore conceptually appropriate to understand why ensembles of models may have generalization abilities surpassing those of the best single models \cite{Hoeting1999, Dietterich2000, Sollich1997, Gal2016, Saatchi2017, Jazbec2025, Samplawski2025}.

We hereafter illustrate our approach on spherical BMs, for which $\Phi=(\ln \overline{Z^n})/N$ can be computed with the replica method when the embedding dimension $N$ of the data is sent to infinity. Our letter is organized as follows. We first present the phase diagram of ensemble learning when the intrinsic dimension of the data, $D$, is low, and discuss the significance of the different phases in the view of large deviations in disordered systems and random matrices. The validity of our results is confirmed by Monte Carlo sampling of the model ensemble. 
We then show how our expression for $\Phi$ can be used to characterize the optimal learning temperature $T^*$ of model ensembles, in agreement with simulations on deep networks. 
Lastly we explain why our theory is exact when $N\to\infty$ irrespectively of the number $K$ of data points, that is, also when $K\sim N$, provided data lie close enough to a low-dimensional manifold. 

\paragraph*{Model and replica calculation.}
The energy of spherical models \cite{Kac1952, Kosterlitz1976}  is given by $E(\boldsymbol{\sigma};\boldsymbol{J}) = -\frac{1}{2}\sum _{i,j} J_{ij} \sigma_i\sigma_j$, where the $N$ spins variables $\sigma_i$ are real-valued and lie on the sphere $\Ss^{N-1}$ of radius $\sqrt{N}$, {\em i.e.} $\sum_i \sigma_i^2=N$ (Fig.~1). The Gram matrix of the data points, $\boldsymbol{C}$, has entries $C_{k\ell}= \frac 1N \boldsymbol{\xi}^k \cdot \boldsymbol{\xi}^\ell$; its eigenvalues, ranked in decreasing order, and eigenvectors are denoted, respectively, as $\chi_k$ and $\boldsymbol{u}_k$ for $k=1,\dots,K$. We choose a Gaussian prior $P_G(\boldsymbol{J})\propto\exp(-\frac{N\gamma}{4} \sum_{i,j} J_{ij}^2)$, where $\gamma$ is a regularization control (hyper)parameter.

In the large-$N$ limit, the intensive logarithm of the average replicated partition function is given by the extremum of 
\begin{eqnarray}
\Phi &=&\frac{1}{4\gamma T}\sum _k (\chi_k)^2 +
\frac{T}{4\gamma}\tr \boldsymbol{Q}^{2} +
\frac{1}{2\gamma}\sum_{k} \chi_k \, |\boldsymbol{M}_k|^2\nonumber \\
&+&\frac 12\log \, \det \left(\boldsymbol {Q}-\sum_k (\boldsymbol{M}_k)(\boldsymbol{M}_k) ^{\top}\right)\  
\label{eq:free-energy}
\end{eqnarray}
over the order parameters characterizing the replicated spin configurations $\boldsymbol{\sigma}_a$, $a=1,\dots,n$: their pairwise overlaps $Q_{ab}$ and their projections $(M_{a})_k$ onto the  eigenmodes $\boldsymbol{u}_k$, see SM Sec.~A. These overlaps and projections describe the expected statistics of data configurations generated by a model randomly drawn from Eq.~\eqref{eq:post}. 


Inspection of the saddle-point $\partial {_{\boldsymbol{Q}}\Phi}\!=\!\partial {_{\boldsymbol{M}}\Phi}\!=\!0$ reveals that, see SM Sec.~B:

\vskip .1cm
\noindent (i) Nonzero vectors $\boldsymbol{M}_k$ are eigenvectors of the overlap matrix $\boldsymbol{Q}$. Let $d \le K$ be the number of these vectors. We call $\nu_k$ and $m_k^2$, respectively, their eigenvalues and squared norms, with $k=1,\dots,d$. 

\vskip .1cm
\noindent (ii) For the restriction of the overlap matrix $\boldsymbol{Q}$ to the orthogonal subspace, of dimension $n-d$, we seek a replica symmetric (RS) solution: $Q_{ab}= \widetilde q \,\delta_{a,b} + q $. 

\vskip .1cm
\noindent (iii) The spherical constraint imposes a compatibility condition between the longitudinal and tranverse subspaces above: $\sum_{k=1}^d \nu_k + (n-d)\, (\widetilde q+q)= n$. 

\begin{figure}[h]
    \centering
    \includegraphics[width=0.8\linewidth]{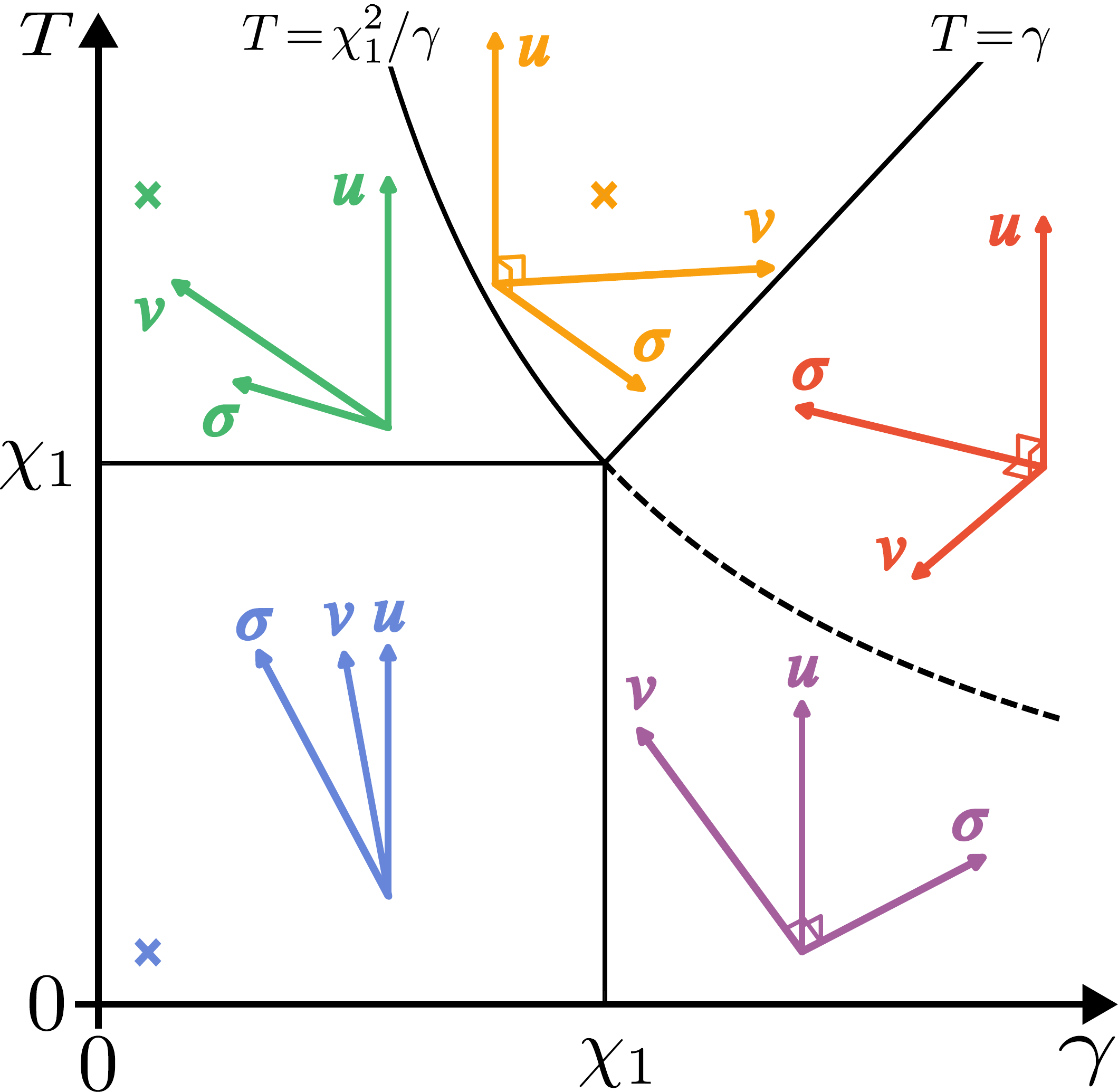}
    \caption{Phase diagram of ensemble learning in the $(\gamma,T)$ plane for one-dimensional training data.
    \textit{Red phase}: learning completely fails, as $\boldsymbol{u}$,  $\boldsymbol{\sigma}$ and $\boldsymbol{v}$ are orthogonal in pairs. \textit{Purple phase}: $\boldsymbol{u}$ is now partially aligned along $\boldsymbol{v}$, but $\boldsymbol{\sigma}$ is orthogonal to both. \textit{Orange phase}: $\boldsymbol{\sigma}$ is partially aligned along $\boldsymbol{v}$, but both are orthogonal to $\boldsymbol{u}$. In the \textit{Blue} and \textit{Green} phases, learning is effective, as $\boldsymbol{u}$, $\boldsymbol{\sigma}$, and $\boldsymbol{v}$ have nonzero dot products. In the \textit{Green} phase, the free energy is frozen to its top feasible value and $|\boldsymbol{v}\cdot\boldsymbol{\sigma}|>|\boldsymbol{v}\cdot\boldsymbol{u}|$, while it is lower and $|\boldsymbol{v}\cdot\boldsymbol{\sigma}|<|\boldsymbol{v}\cdot\boldsymbol{u}|$ in the \textit{Blue} phase. 
    Crosses $\times$ locate the points for which the large-deviation rate functions of $f$ are shown in Fig.~2.}
    \label{fig:1}
\end{figure}

\vskip .2cm
These RS equations fully determine the parameters $\{\nu_k,m_k^2\}$ and the  overlaps $\widetilde q,q$ \footnote{Our solution is stable against longitudinal and transverse fluctuations in the replica space (SM Sec.~C).} provided the number of replicas is larger than 
\begin{equation}\label{n_c}
    n_c= - \sum_{k(\le d)} \frac{\chi_k-\sqrt{\gamma\, T}}{T-\sqrt{\gamma\, T}} \ , 
\end{equation}
when $n_c<0$. As $n$ is lowered below $n_c$, the solution remains frozen, and ${\Phi}$ linearly depends on $n$, {\em i.e.} ${\Phi}(n)={\Phi}(n_c)+(n-n_c){\Phi}'(n_c)$  \cite{Pastore2019}. 

The physical origin of this freezing is  the following. Training (under a given prior) amounts to searching for models $\boldsymbol{J}$ with  high free energies $f=-\frac 1N \log Z(\boldsymbol{J})$, while maintaining the energy of the data low, see Eq.~\eqref{eq:post}. The negative  number $n$ of replicas acts as a thermodynamical force pushing $f$ well above its typical value with the prior $\rho$ (corresponding to $n=0$). These large deviations are characterized by the rate function $\mathcal{I}(f)=-\frac 1N \log \text{Proba}(f)$, which is the Legendre transform of $\Phi(n)$, see SM Sec.~D. 
In practice, ${\cal I}$ is infinite when $f$ is greater than some $f_c$, as higher free energies are extremely unlikely, {\em i.e.} have $e^{-{O}(N^2)}$ rather than $e^{-{O}(N)}$ probabilities \cite{Dean2006}. Freezing takes place when the right edge of ${\cal I}$ is hit, {\em i.e.} when $f_c=-\Phi'(n_c)$, or, equivalently, $n_c=-{\cal I}'(f_c)$ due to Legendre relations.

\paragraph*{One-dimensional case.}
We first consider $K$ data points aligned along the direction $\boldsymbol{u}\,( =\boldsymbol{u}_1)$, {\em i.e.} $D=1$ nonzero eigenvalue $\chi_1$, equal to $K$ if the points have average squared norm $N$, see Fig.~1. The other directions of interest for a representative sampled model $\boldsymbol{J}$ are: the generated configurations $\boldsymbol{\sigma}$, and the ground state $\boldsymbol{v}$, {\em i.e.} the top eigenvector of $\boldsymbol{J}$. To characterize $\boldsymbol{v}$, we  extend the replica approach described above by including $n'\to 0$ replicas thermalized at zero temperature in addition to the $n=-K/T$ already present replicas, see SM Sec.~D.  

Our replica calculation shows that the pairwise overlaps between  $\boldsymbol{u}$, $\boldsymbol{v}$, and $\boldsymbol{\sigma}$ are nonzero for small enough regularization $\gamma$ only, see Fig.~1. Overfitting is present for low enough $\gamma$ and $T$ {\em i.e.} in the blue phase: the log-likelihood $\ell=-E-\log Z$ of generated data,  $\ell_{\text{gen}}$, is lower than the one of training data, $\ell_{\text{train}}$ (SM Sec. E).  The two log-likelihoods become equal on the $T=\chi_1$ line separating the blue and green phases and identifying the freezing transition mentioned above, see rate functions in Fig.~2. In the green phase,  $\ell_{\text{gen}} > \ell_{\text{train}}$:  the overlap between $\boldsymbol{\sigma}$ and $\boldsymbol{v}$ is large, but both gradually decorrelate with the data direction $\boldsymbol{u}$ as $T$ grows. Ensemble training at an appropriate temperature $T$ thus helps prevent overfitting, as we will confirm later.

\begin{figure}[h]
    \centering
    \includegraphics[width=1.\linewidth]{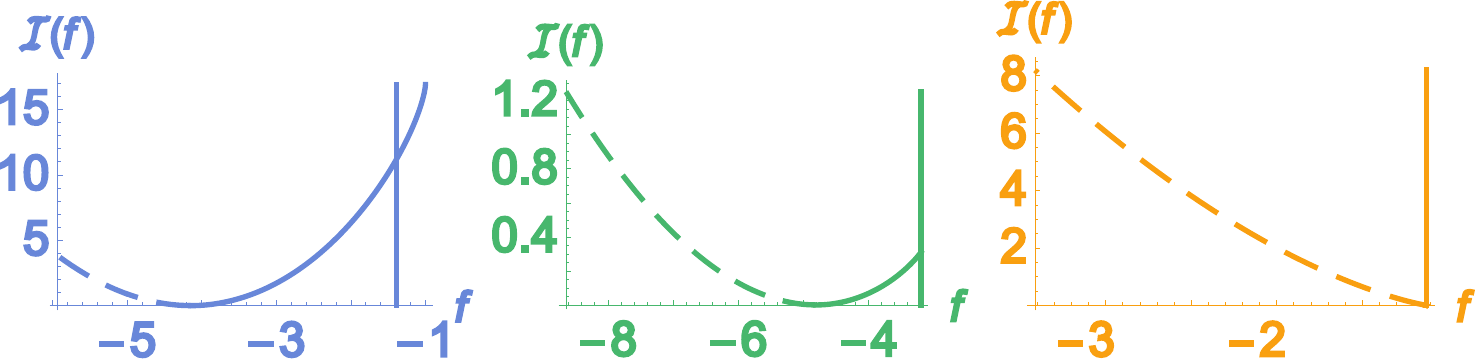}
    \caption{Rate functions $\mathcal{I}$ for the free energy $f$ under the prior $\rho$ for the three points marked with $\times$ in Fig.~1; Vertical bars locate the typical free energies $f^*$ of models sampled from $P_T$. Deviations of $f$ below its typical value under the prior $\rho$, corresponding to $n>0$, are indicated with dashes.
    \textit{Blue phase:} $f^*$ is such  that ${\cal I}'(f^*)=-n$, and is larger than its typical value under the prior $\rho$, corresponding to $\mathcal{I}'=0$.  \textit{Green phase:} $f^*$ is frozen at the right edge $f_c$ of the rate function, with slope $-n_c$, see Eq.~\eqref{n_c}.  \textit{Orange phase:} same as green with $n_c=0$: typical free energies under $P_T$ and $\rho$ coincide.  }
    \label{fig:placeholder}
\end{figure}

\begin{figure}[h]
    \centering
    \includegraphics[width=1.0\linewidth]{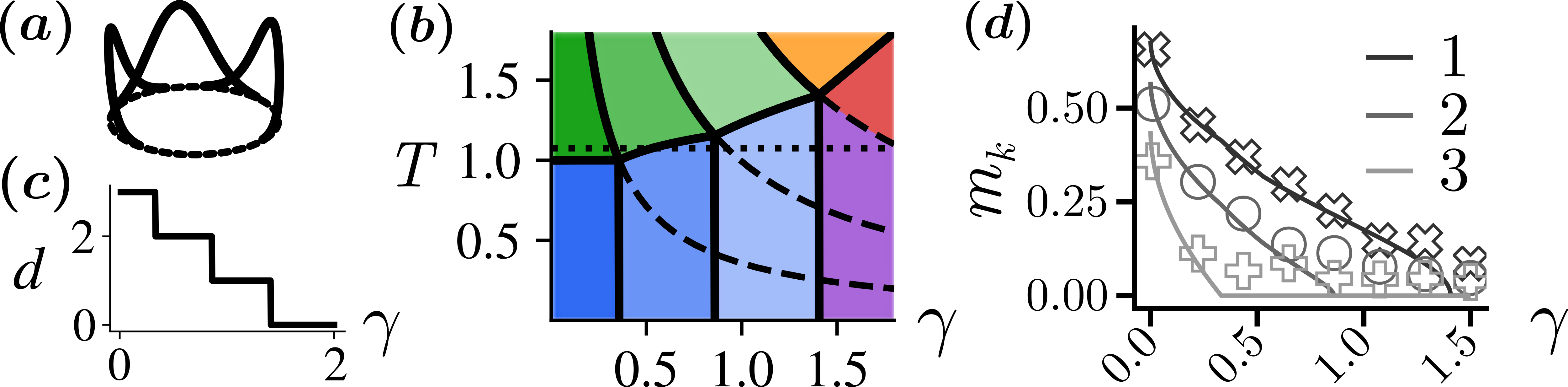}
    \caption{Cascade of phase transitions for $K=3$ data points. $\boldsymbol{(a)}$ The training data are three  normalized bumps $\xi^k_i\propto e^{-(i-i^k)^2/(2\tau^2)}$, centered around random $i^k$'s, with additive Gaussian noise of variance $10^{-6}$; parameters are $N=1000$, $\tau=100$. $\boldsymbol{(b)}$ Phase diagram of ensemble learning. Note the presence of three blue and green phases, depending on the number $d=0,1,2,3$ of non-zero magnetizations.$\boldsymbol{(c)}$ $d$ vs. $\gamma$ and $\boldsymbol{(d)}$ $m_k$ vs. $\gamma$   for $T=1.075$  (black dotted line in $\boldsymbol{(b)}$); dots show MC estimates (with small error bars) of $m_k^2$ obtained  by projecting the generated data covariance matrix along $\boldsymbol{u}_k$.} 
    \label{fig:langevin-simulations}
\end{figure}

\paragraph*{Multidimensional case and the cascade phenomenon.}
The phase diagram in Fig.~1 was derived for $D=1$, regardless of the number $K$ of data points, which affects only the amplitude of the unique nonzero eigenvalue $\chi_1$.
As a completely different example, we  now consider $K$ generically distributed data points, {\em i.e.} such that all eigenvalues $\chi_k$, $k=1,\dots,K$, are nonzero. Replica theory predicts a cascade of phase transitions as $\gamma$ is decreased at fixed $T$, with more and more nonzero magnetizations $m_k$. An illustration in the case of $K=3$ bump-like data on a ring is shown in Fig. \ref{fig:langevin-simulations}(a,b,c). Locations of the transition lines depend on the eigenvalues $\chi_k$'s, see SM Sec.~B. The cascade phenomenon and, more generally, the existence of various learning phases can be confirmed using rigorous results on the large deviations  of the top eigenvalues and eigenvectors of Gaussian random matrices with finite-rank perturbations, see \cite{Maida2019} and SM Sec.~D for $K=1$, and \cite{Guionnet2021,Jorge} for $K\ge 2$.

To assess the validity of the results for finite $N$, we implement Monte Carlo (MC) sampling of the posterior distribution $P_T$ in Eq.~\eqref{eq:post} over the space of real symmetric matrices $\boldsymbol{J}$. MC moves are proposed with the Euler–Maruyama discretization of the overdamped Langevin dynamics
\begin{align}
    \mathrm{d}\boldsymbol{J}_t=\partial_{\boldsymbol{J}}\log P_1(\boldsymbol{J}\vert\mathcal{D})\,\mathrm{d}t+\sqrt{2T}\,\mathrm{d}\boldsymbol{W}_t,
    \label{eq:langevin}
\end{align}where $\partial_{\boldsymbol{J}} \log P_1(\boldsymbol{J}\vert\mathcal{D})=\frac{1}{2}\left(\boldsymbol{C}- K \langle \boldsymbol{\sigma} \cdot \boldsymbol{\sigma}^\top\rangle_t\right)-\frac{N\gamma}{2}\boldsymbol{J}_t$, and $\mathrm{d}\boldsymbol{W}$ is a collection of  independent Brownians (up to symmetry). MC moves are then accepted or rejected with Metropolis, see SM Sec.~F and \cite{Besag1994, Roberts1996}. MC estimates of the magnetizations $m_k$'s are in very good agreement with  theoretical predictions for the bump-like data, see Fig. \ref{fig:langevin-simulations}(d).

\paragraph*{Optimal ensemble of models.} 
Let us define the posterior predictive at temperature $T$, $\text{PP}_T(\boldsymbol{\sigma}|{\cal D})$, as the probability that the ensemble of models $P_T(\boldsymbol{J}\vert\mathcal{D})$ generate a data point $\boldsymbol{\sigma}$ \cite{Roberts1965}. Good choices of hyperparameters should guarantee that test data points $\boldsymbol{\xi}'$, not included in the training data ${\cal D}$, are assigned large $\text{PP}_T$ values. The posterior predictive on test data can be readily computed from the knowledge of ${\Phi}$ through
\begin{align}\label{eq:pp}
    \frac 1N \ln \text{PP}_T(\boldsymbol{\xi}'\vert\mathcal{D}) =\Phi(\boldsymbol{\xi}',\mathcal{D})-\Phi(\mathcal{D})\,.
\end{align}
Here, ${\Phi}({\cal D})$ is the result of the optimization of Eq. \eqref{eq:free-energy},  while $\Phi(\boldsymbol{\xi}',\mathcal{D})$ is evaluated with $n=-K/T-1$ replicas and the overlaps of the $K$ training points in ${\cal D}$ with $\sqrt{T} \, \boldsymbol{\xi}'$ added to $\boldsymbol{C}$, see SM Sec. E. We then define the cross entropy, $\mathrm{CE}(T)$, as minus the average  of $\frac 1N\ln \mathrm{PP}_T$ over $\mathcal{D}$ and $\boldsymbol{\xi}'$. The learning temperature $T^\star$  minimizing $\mathrm{CE}(T)$ identifies the optimal ensemble of models. The efficient computation of the log posterior predictive  in Eq.~\eqref{eq:pp} allows us to estimate $T^*$ for complex data distributions. We hereafter consider  CIFAR-10, which consists of color images of size $N=3\times 32\times32=3072$. 

The optimal temperature $T^\star$ grows with the dissimilarity between the training and test data, as intuitively expected. This is shown for $K=1$ training point $\boldsymbol{\xi}^1$ and 1 test point $\boldsymbol{\xi}'$ with varying degrees of similarity  (Fig.~4(a)), and for $K=200$ images from CIFAR-10 (Fig.~4(b)). In that case, we compare the training and test data through the cosine similarity of their representations in a deep pretrained model suited for image classification, see SM Sec.~F. The existence of an optimal $0<T^*<1$ is quite robust. 
As an illustration, following \cite{Wenzel2020}, we use stochastic gradient Langevin dynamics \cite{Welling2011} to sample ensembles of deep convolutional neural networks \cite{He2015} of about $3\times 10^5$ parameters for CIFAR-10 image classification at varying learning temperatures, see Fig ~4(c). Here, we take $\text{PP}_T$ as the conditional probability that the ensemble assigns, given an image, to the different classes. Ensembles corresponding to $T^*$ perform better on outlier data  than MAP ($T=0$) and Bayes posterior ($T=1$), consistently with empirical evidence in spite of stark differences between the learning task, the model architecture, and the sampling algorithm \cite{Wenzel2020}. 

\begin{figure}[h]
    \centering
    \includegraphics[width=1\linewidth]{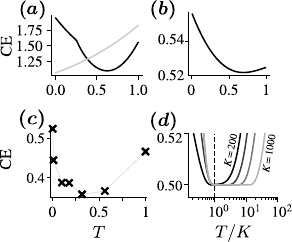}
    \caption{Cross entropy curves. $\text{CE}(T)$ computed from Eq. \eqref{eq:pp} for $\boldsymbol{(a)}$ $K=1$ training and test points with $\gamma=0.6$, cosine similarity $=0.9$ (gray) and  $\gamma=0.1$, cosine similarity $=0.1$ (black).  $\boldsymbol{(b)}$ $\text{CE}(T)$ for the spherical BMs and $K=200$ CIFAR-10 training images, 1 test image with average cosine similarity $\simeq 0.61$; $\gamma=0.001$. $\boldsymbol{(c)}$ $\text{CE}(T)$ estimated with MC for an ensemble of deep classifiers 
    ResNet-20s, with $K=50,000$ and the $1,000$ farthest test points; $\gamma=0.01$. $\boldsymbol{(d)}$ $\text{CE}$ (solid lines) vs. $T/K$ for $K=200,300,500,1000$ training points along a line (as in Fig.~1), and a test point with cosine similarity $=0.8$ with the line direction; dashed vertical line: $T_c/K\sim 1$.
    }
    \label{fig:CE}
\end{figure}

\paragraph*{Nearly finite-dimensional data.} To characterize the optimal temperature $T^*$ further, we now concentrate on the case of data living in a thin $D$-dimensional slab, with $D$ finite compared to $K$. More precisely, we assume that, as $K$ grows, a finite number $D$ of eigenvalues $\chi_k$ scale linearly with $K$, while the remaining $K-D$ eigenvalues are much smaller, {\em e.g.} bounded or sublinear. 

Optimization of $\Phi$ in Eq.~\eqref{eq:free-energy} when $K\gg 1$ yields, under conditions over $\gamma, T$ specified hereafter,  that $\boldsymbol{v}_k\simeq \boldsymbol{u}_k$ and $m_k^2\simeq \chi_k/K$ for $k=1,\dots,D$.  
Generated data $\boldsymbol{\sigma}$ have  vanishing projections along the other directions as long as the regularization is strong enough, {\em i.e.}, $\gamma>\gamma_c$ where
\begin{equation}
\gamma_c=\chi_{D+1}\delta_K+\frac{D}{K}\chi_{D+1}^2,\ \text{with} \ \ \delta_K= 1 - \frac 1K \sum_{k=1}^D \chi_k \,.
\end{equation}
For data $\boldsymbol{\xi}$ lying on average on $\mathbb{S}^{N-1}$, $\delta_K$ is comprised between 0 and 1 and is, in the language of statistics, the fraction of variance unexplained by the $D$ projections of the data along the $\boldsymbol{v}_k$'s. The data model in Fig.~1 corresponds to the simplest possible case: $D=1$, $\delta_K=0$. 

When $\gamma> \gamma_c$, the temperature $T_c$ at which freezing occurs, {\em i.e.} separating the blue and green phases in Fig.~1 depends on the scaling of $\delta_K$, see SM Sec.~I:

\vskip .1cm
\noindent {\em (i) Thick slab, i.e. strong transverse fluctuations $\delta _K  \gg K^{-1/2}$.} One has $T_c\sim \gamma/\delta_K^2$. The squared projections of generated data along the $K-D$ transverse directions sum up to $\widetilde{q}\sim \delta_K$. 

\vskip .1cm
\noindent {\em (ii) Thin slab, i.e. weak transverse fluctuations $|\delta _K |  \lesssim K^{-1/2}$.} In this regime, $T_c\sim K/D$ and  $\widetilde{q}\sim\sqrt{D\gamma/K}$. 

\vskip .1cm
Figure \ref{fig:CE}(d) shows the cross entropy CE for the one-dimensional setting of Fig.~1 and one outlier test point. We observe that, as the temperature is raised from $T=0$, CE decreases and reaches its minimum at $T^*$ scaling linearly with $K$, and approximately given by $T_c$, see regime \textit{(ii)} above. This confirms that sampling an ensemble of models at a finite temperature reduces overfitting compared to MAP.

The expressions of $T_c$ above were derived with the replica method when $K\gg1$, after having first sent $N\to\infty$. We now examine their validity, within our nearly finite-dimensional data framework, when $N$ is finite, of the order of $K$. Figure~5(a) reports MC simulations and replica predictions for the same  bump-like data on a ring as in Fig.~3(a), but for a much larger number of data points, $K=800$, comparable to the dimension,  $N=1000$. A large interval of regularization strengths $\gamma$  is compatible with $d=2$ nonzero magnetizations $m_1,m_2$.
These bump-like data, indeed, naturally lie very close to a planar embedding of the ring: $D=2$ eigenvectors $\boldsymbol{u}_1,\boldsymbol{u}_2$, corresponding to cosine and sine waves, have similar eigenvalues $\chi_1\simeq \chi_2$ scaling linearly with $K$, while the other eigenvalues $\chi_{k\ge 3}$ remain small \footnote{The low-dimensional nature of bump-like data is well known in computational neuroscience, see S. Amari. \textit{Biol. Cybern.} 27, 77–87 (1977).}. We observe a striking agreement between theory and simulations \footnote{Additional information about the low-dimensional nature of the data generated by the model $\boldsymbol{J}$ sampled from the posterior $P_T$ can be found in SM Section G.}. 

\begin{figure}[h]
    \centering
    \includegraphics[width=1.0\linewidth]{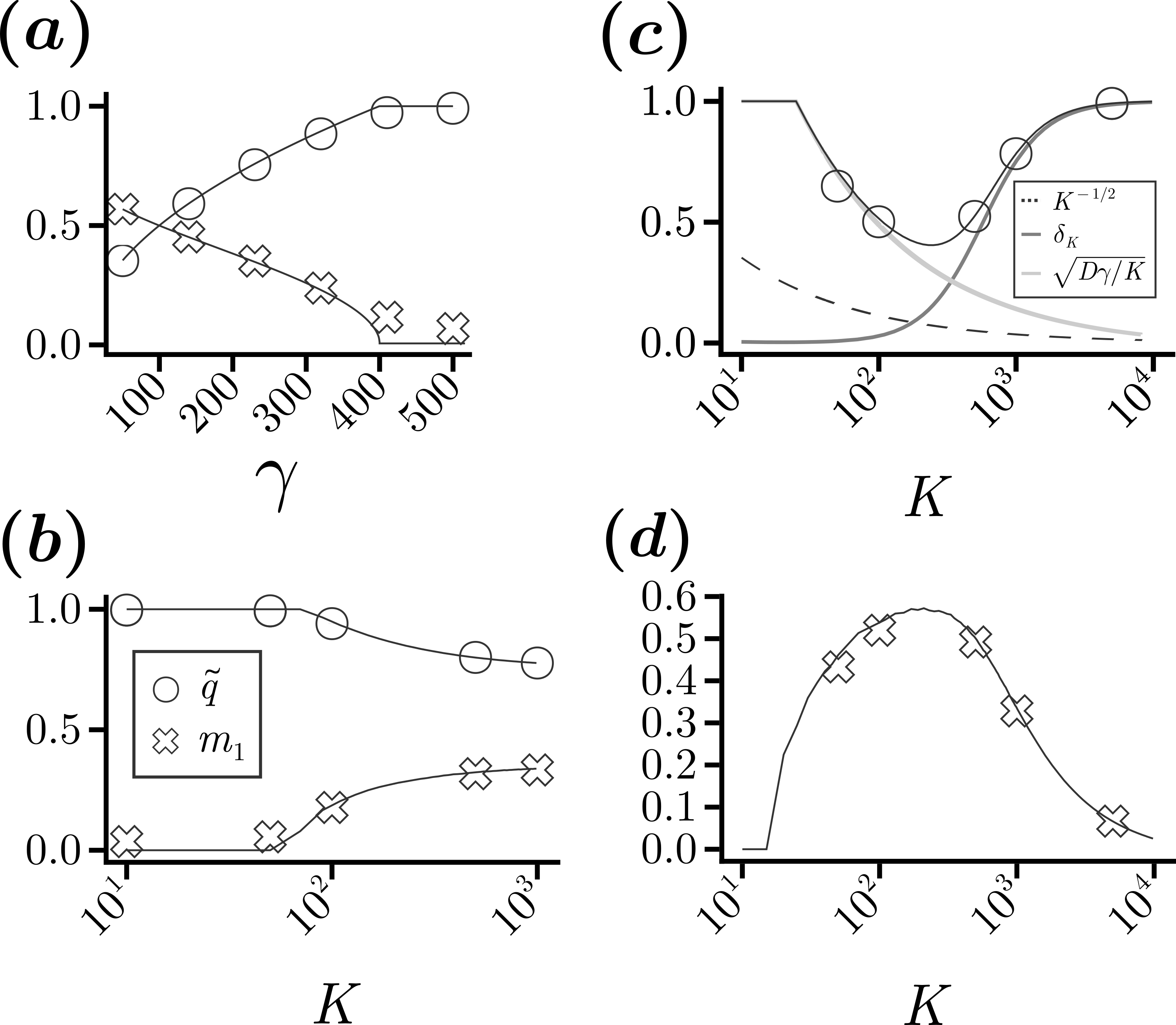}
    \caption{Monte Carlo simulations (symbols) and comparison with replica predictions (lines) for nearly finite-dimensional  data. Crosses: fraction $\widetilde{q}$ of the variance of generated data  outside the $D=2$ plane; circles: root mean squared projection $m_1$ along $\boldsymbol{u}_1$ ($m_2$ is very similar and not shown). Data are bumps of squared norm $N$, width $\tau=0.1N$ as in Fig.~3.  
    $\boldsymbol{(a)}$ Case of $K=800$ noiseless bumps for fixed $N=1000$, $T=2$, and variable $\gamma$.  
    $\boldsymbol{(b)}$ Case of  $K$ noisy bumps as in Fig.~3(a) for fixed $N=1000$, $\gamma=10$, $T=0.5$. 
    $\boldsymbol{(c),(d)}$ Case of variable $K$ and $N$ at fixed ratio $K/N=0.8$; $\gamma=10$, $T=0.5$. The crossover between regimes \textit{(i)} and \textit{(ii)} is visible for $\widetilde q$ in the top panel.}
    \label{fig:cann}
\end{figure}

 To understand why this is so, let us consider again our theoretical predictions for nearly finite-dimensional data (derived for finite $K$ and $N\to \infty$). When $T \ll T_c$, neither the $m_k$'s nor $\widetilde q$ depend on the value of $T$, and the system is effectively in a zero-temperature regime, in spite of the fact that $T$ is finite and might even be large. This is confirmed by a direct analysis of the MAP interaction matrix, which maximizes the posterior in Eq.~\eqref{eq:post}. Calculation of ${\boldsymbol J}_\text{MAP}$ can be carried out without replicas, establishing an explicit connection with random matrix theory, see SM Sec.~H. Irrespective of whether $K/N$ is small  or large, the expressions for $\widetilde{q}$ and the $m_k^2$'s are the same, in regimes \textit{(i)} and \textit{(ii)}, as those obtained with replicas under the assumption $K$ finite, $N\to \infty$ as long as $T<T_c$. The dataset  considered in Fig.~5(a)  is in this regime as $\delta_{800}\simeq 0.0008 \ll(800)^{-1/2} \simeq 0.035$, $\gamma \gg \gamma_c\simeq 0.00005$, and $T=2 \ll T_c\simeq 400$. MC simulations provide further confirmation of the accuracy of our replica results, when $K$ grows at fixed $N$ (Fig.~\ref{fig:cann}(b)) and when both $K$ and $N$ increase at a fixed ratio (Fig.~\ref{fig:cann}(c,d)).  In this setting, replica calculations can be carried out, in spite of the fact the replica number $n$ is proportional to $N$,  see SM Sec.~J.

\paragraph*{Perspectives.} In this Letter, we showed how ensemble learning can be analytically studied with the replica method due to a duality between finite-temperature training of high-dimensional energy-based models and large deviations of the free energy of spin-glass models. Interestingly, when data are of nearly finite dimension and the training temperature not too high, the replica theory can be worked out for arbitrarily large number of data points compared to the  embedding dimension, see SM Sec.~J. This is in striking contrast with the difficulty in deriving large deviations of the free energy for spin-glass models with unstructured disorder \cite{Parisi2010}. It would be very interesting to see if our approach can be applied to derive the learning curves of binary BMs with nearly low-dimensional data, for which random matrix theory is inoperant. Such a study could  be relevant for applications, \emph{e.g.}, to help to understand how to optimally choose the temperature for generating data \cite{Russ2020}. Our calculation can, indeed, be easily modified to compare data sampled at different temperatures, SM Sec.~D and \cite{Nishimori2024}.

The replica formalism presented herein could also apply to models that achieve greater expressivity through latent variables, such as the Restricted Boltzmann Machine (RBM) \cite{Decelle2021}, see also \cite{FernandezdeCossioDiaz2022, Fachechi2024}. Sparse priors, for instance, are known to induce compositional representations in random ensembles of RBMs, corresponding to learning temperature $T=\infty$ \cite{Tubiana2017}. How the distribution of these representations adapt to finite $T$ and depend on the features of the data remains a largely open question.

\section*{Acknowledgments} We are thankful to M. Pastore and P. Urbani for useful discussions, and to M. Pastore for reading the manuscript. We acknowledge financial support from the ANR grant MEMNET.

\newpage
\bibliographystyle{apsrev4-2}
\bibliography{refs}

\end{document}


\title{Supplemental Material:\\
\textit{Replica Theory of Spherical Boltzmann Machine Ensembles}}

\author{Thomas Tulinski}
\affiliation{Laboratoire de Physique de l'\'Ecole Normale Sup\'erieure, PSL, CNRS UMR8023, Sorbonne Universit\'e, 24 rue Lhomond, 75005 Paris, France}
\author{Jorge Fernandez-de-Cossio-Diaz}
\affiliation{Institut de Physique Th\'eorique, Universit\'e Paris-Saclay, CNRS UMR3681, CEA, Gif-sur-Yvette, France}
\author{Simona Cocco}
\affiliation{Laboratoire de Physique de l'\'Ecole Normale Sup\'erieure, PSL, CNRS UMR8023, Sorbonne Universit\'e, 24 rue Lhomond, 75005 Paris, France}
\author{R\'emi Monasson}
\affiliation{Laboratoire de Physique de l'\'Ecole Normale Sup\'erieure, PSL, CNRS UMR8023, Sorbonne Universit\'e, 24 rue Lhomond, 75005 Paris, France}

\date{\today}

\maketitle

\tableofcontents

\newpage
\section{Notations}
\begin{table}[h]
    \caption{\textbf{List of important parameters}}
    \label{tab:parameters}
    \begin{ruledtabular}
    \begin{tabular}{cl}
        Notation & Meaning \\
        \hline
        $n$                             & Replica number\\
        $N$                             & Embedding dimension of the data\\
        $D$                             & Intrinsic dimension of the data\\       
        $K$                             & Number of data\\
        $T$                             & Learning temperature\\
        $T'$                          & Generative temperature\\
        $\gamma$                        & L2 regularization strength\\
    \end{tabular}
    \end{ruledtabular}
\end{table}

\section{A. Replica calculation of the marginal likelihood}\label{derivation-SCGF}
Here, we are interested in the quantity 
\begin{align}\label{eq:marginal-likelihood}
    Y(\mathcal{D})\,\,
    &=\lim_{n\to-K/T}\int\mathrm{d}\boldsymbol{J}\left[P_G(\boldsymbol{J})\,\prod_{k=1}^K\frac{e^{-E(\boldsymbol{\xi}_k;\boldsymbol{J})/T'}}{Z(\boldsymbol{J})}\right]^{\frac{1}{T}}\nonumber\\
    &=\int\mathrm{d}\boldsymbol{J}\,P_G(\boldsymbol{J})^\frac{1}{T}\,e^{-\sum_{k=1}^KE(\boldsymbol{\xi}_k;\boldsymbol{J})/T'T}\;Z(\boldsymbol{J})^{-\frac{K}{T}}\nonumber\\
    &=\lim_{n\to-K/T}\int\mathrm{d}\boldsymbol{J}\,\rho_J(\boldsymbol{J})\;Z(\boldsymbol{J})^n\nonumber\\
    &=\lim_{n\to-K/T}\overline{Z(\boldsymbol{J})^n}
\end{align}
for spherical Boltzmann machines, $E(\boldsymbol{\sigma};\boldsymbol{J})=-\frac{1}{2}\boldsymbol{\sigma}^\top\boldsymbol{J}\boldsymbol{\sigma}$, $\Vert\boldsymbol{\sigma}\Vert^2=N$, and Wigner priors, $P_G(\boldsymbol{J})\propto e^{-\frac{N\gamma}{4}\mathrm{tr}(\boldsymbol{J}^2)}$, $\boldsymbol{J}=\boldsymbol{J}^\top$, where $\gamma$ controls the strength of the L2 regularization and $T$ the training temperature. Unless $T=1$, the distribution $P_T(\boldsymbol{J}\vert\mathcal{D})$ defined by $Y(D)$ does not coincide with a Bayesian posterior. The reason that $T$ cannot be absorbed by rescaling the hyperparameter $T'$ is $Z(\boldsymbol{J})^{1/T}\neq Z(\boldsymbol{J}/T)$. We set $T'=1$ over the course of the derivation and can recover the dependence at any point by rescaling $\gamma\mapsto\gamma T'^2,$ e.g. to characterize the ground state in Sec.~D.

We begin by computing \begin{align}\label{eq:CGF}
    \Phi(n)=\lim_{N\to\infty}\frac{1}{N}\ln\overline{Z(\boldsymbol{J})^n}
\end{align}
holding $K$ fixed, and then sending $K$ to infinity. 

The starting point is the introduction of $n$ replicas $\{\boldsymbol{\sigma}^a\}_{a=1}^n$ to form the replicated partition function \begin{align}
    Z(\boldsymbol{J})^n=\int\prod_{a=1}^n\mathrm{d}\boldsymbol{\sigma}^a\,\delta(\Vert\boldsymbol{\sigma}^a\Vert^2-N)\,e^{\frac{1}{2}\sum_{a=1}^n\boldsymbol{\sigma}^{a\top}\boldsymbol{J}\boldsymbol{\sigma}^a},\nonumber
\end{align}
which we want to average over $\boldsymbol{J}$. For that, we must compute $N$ independent Gaussian integrals $i$ coming from the diagonal of $\boldsymbol{J}$ and, by symmetry, an additional $\frac{1}{2}N(N-1)$ independent Gaussian integrals $i<j$ from the off-diagonal. 
Carrying out the Gaussian integrals,\begin{align}
    \overline{Z(\boldsymbol{J})^n}&\propto\int\prod_{a=1}^n\mathrm{d}\boldsymbol{\sigma}^a\;\prod_{a=1}^n\delta(\Vert\boldsymbol{\sigma}^a\Vert^2-N)\prod_{i=1}^Ne^{\frac{T}{N\gamma}\left(\frac{1}{2T}\sum_{k=1}^K(\xi_i^k)^2+\frac{1}{2}\sum_{a=1}^n(\sigma_i^a)^2\right)^2}\prod_{1\le i<j\le N}e^{\frac{T}{2N\gamma}\left(\frac{1}{T}\sum_{k=1}^K\xi_i^k\xi_j^k+\sum_{a=1}^n\sigma_i^a\sigma_j^a\right)^2}\nonumber\\
    &\propto\int\prod_{a=1}^n\mathrm{d}\boldsymbol{\sigma}^a\;\prod_{a=1}^n\delta(\Vert\boldsymbol{\sigma}^a\Vert^2-N)\;e^{\frac{T}{4N\gamma}\sum_{i,j=1}^N\left(\frac{1}{T}\sum_{k=1}^K\xi_i^k\xi_j^k+\sum_{a=1}^n\sigma_i^a\sigma_j^a\right)^2}.\nonumber
\end{align}
It is convenient to work in the basis where $C_{ij}=\frac{1}{K}\sum_{k=1}^K\xi_i^k\xi_j^k$ is diagonal, whose spectral decomposition we write $\boldsymbol{C}=\frac{1}{K}\sum_{k=1}^K\chi_k\boldsymbol{u}_k\boldsymbol{u}_k^\top$ where $\boldsymbol{D}_\chi={\rm diag}(\chi_k)_{k=1}^K$ are the eigenvalues of $X_{kl}=\frac{1}{N}\boldsymbol{\xi}_k\cdot\boldsymbol{\xi}_l$ and $\frac{\boldsymbol{u}_k}{\sqrt{N}}$ are orthonormal eigenvectors of $\boldsymbol{C}$. We introduce the overlap matrices $Q_{ab}=\frac{1}{N}\boldsymbol{\sigma}_a^\top\boldsymbol{\sigma}_b$ and $M_{ak}=\frac{1}{N}\boldsymbol{\sigma}_a^\top\boldsymbol{u}_k$, in terms of which 
\begin{align}
    \overline{Z(\boldsymbol{J})^n}
    \propto e^{\frac{N}{4\gamma T}\mathrm{tr}(\boldsymbol{X}^2)}\int\prod_{a=1}^n\mathrm{d}\boldsymbol{\sigma}^a\;\prod_{a=1}^n\delta(\Vert\boldsymbol{\sigma}^a\Vert^2-N)\;e^{\frac{NT}{4\gamma}\mathrm{tr}(\boldsymbol{Q}^2)+\frac{N}{2\gamma}\mathrm{tr}(\boldsymbol{M}\boldsymbol{D}_\chi\boldsymbol{M}^\top)}.\nonumber
\end{align}
To compute the entropy of configurations for any given value of the overlaps, $\boldsymbol{Q},\boldsymbol{M}$, we make use of the identities\begin{align}
    \prod_{a=1}^n \delta\!\left(NQ_{aa}-\sum_{i=1}^N(\sigma_i^a)^2\right)&\propto\int\prod_{a=1}^n \mathrm{d}\widehat Q_{aa}\;e^{-\frac{N}{2}\sum_{a=1}^n \widehat Q_{aa}\left(Q_{aa}-\frac{1}{N}\sum_{i=1}^N(\sigma_i^a)^2\right)},\nonumber\\    
    1=\int\prod_{1\le a<b\le n} \mathrm{d}Q_{ab}\;\delta\left(NQ_{ab}-\sum_{i=1}^N\sigma_i^a\sigma_i^b\right)&\propto\int\prod_{1\le a<b\le n} \mathrm{d}Q_{ab}\;\int\prod_{1\le a<b\le n}\mathrm{d}\widehat Q_{ab}\;e^{-N\sum_{1\le a<b\le n}\widehat Q_{ab}\left(Q_{ab}-\frac{1}{N}\sum_{i=1}^N \sigma_i^a\sigma_i^b\right)},\nonumber\\
    1=\int\prod_{a=1}^n\prod_{k=1}^K\mathrm{d}M_{ak}\;\delta\left(M_{ak}-\tfrac{1}{N}\sum_{i=1}^N\sigma_i^au_i^k\right)&\propto\int\prod_{a=1}^n\prod_{k=1}^K\mathrm{d}M_{ak}\int\prod_{a=1}^n\prod_{k=1}^K\mathrm{d}\widehat{M}_{ak}\;e^{-N\sum_{a=1}^n\sum_{k=1}^K\widehat{M}_{ak}\left(M_{ak}-\frac{1}{N}\sum_{i=1}^N\sigma_{i}^au_{i}^k\right)}.\nonumber
\end{align}
Writing $\mathrm{d}(\boldsymbol{Q},\widehat{\boldsymbol{Q}},\boldsymbol{M},\widehat{\boldsymbol{M}})=\prod_{a<b}\mathrm{d}Q_{ab}\prod_{a<b}\frac{N}{2\pi i}\mathrm{d}\widehat{Q}_{ab}\prod_a\frac{N}{4\pi i}\mathrm{d}\widehat{Q}_{aa}\!\prod_{ak}\mathrm{d}M_{ak}\prod_{ak}\frac{N}{2\pi i}\mathrm{d}\widehat{M}_{ak}$, one has 
\begin{align}
\overline{Z(\boldsymbol{J})^n}&\propto\int\mathrm{d}(\boldsymbol{Q},\widehat{\boldsymbol{Q}},\boldsymbol{M},\widehat{\boldsymbol{M}})\;e^{\,N \Phi(\boldsymbol{Q},\widehat{\boldsymbol{Q}},\boldsymbol{M},\widehat{\boldsymbol{M}})}\nonumber\\
g(\boldsymbol{Q},\widehat{\boldsymbol{Q}},\boldsymbol{M},\widehat{\boldsymbol{M}})&=\frac{1}{4\gamma T}\mathrm{tr}(\boldsymbol{X}^2)+\frac{T}{4\gamma}\mathrm{tr}(\boldsymbol{Q}^2)+\frac{1}{2\gamma}\mathrm{tr}(\boldsymbol{M}\boldsymbol{D}_\chi\boldsymbol{M}^\top)+s(\boldsymbol{Q},\widehat{\boldsymbol{Q}},\boldsymbol{M},\widehat{\boldsymbol{M}})\nonumber\\
s(\boldsymbol{Q},\widehat{\boldsymbol{Q}},\boldsymbol{M},\widehat{\boldsymbol{M}})&=-\frac{1}{2}\mathrm{tr}(\boldsymbol{Q}\widehat{\boldsymbol{Q}})-\mathrm{tr}(\boldsymbol{M}\widehat{\boldsymbol{M}}^\top)+\frac{1}{N}\sum_{i=1}^N\ln z_i(\widehat{\boldsymbol{Q}},\widehat{\boldsymbol{M}})\nonumber\\
z_i(\widehat{\boldsymbol{Q}}, \widehat{\boldsymbol{M}})&=\int\mathrm{d}\boldsymbol{\sigma}_i\; e^{\,\frac{1}{2}\boldsymbol{\sigma}_i^\top \widehat{\boldsymbol{Q}}\boldsymbol{\sigma}_i+\boldsymbol{u}_i\widehat{\boldsymbol{M}}^\top\boldsymbol{\sigma}_i},\nonumber
\end{align}
where $\boldsymbol{\sigma}_i=(\sigma_{i}^a)_{a=1}^n$ and $\boldsymbol{u}_i=(u_{i}^k)_{k=1}^K$. Conditioned on $\widehat{\boldsymbol{Q}}$ and $\widehat{\boldsymbol{M}}$, sites $i\neq j$ are effectively independent, \begin{align}
\langle\sigma_i^a\sigma_j^b\rangle&=\langle\sigma_i^a\rangle\langle\sigma_j^b\rangle,\nonumber
\end{align}but now replicas are  $a\neq b$ coupled. The stationarity conditions $\frac{\partial\mathcal{S}}{\partial\hat{\boldsymbol{Q}}}=\frac{\partial\mathcal{S}}{\partial\hat{\boldsymbol{M}}}=0$ indeed give\begin{align}
    Q_{aa}&=1, & Q_{ab}&=\frac{1}{N}\sum_{i=1}^N\langle\sigma_i^a\sigma_i^b\rangle \quad (a<b), & M_{ak}&=\frac{1}{N}\sum_{i=1}^N\langle\sigma_i^a\rangle u_i^k.
\end{align}
Completing the square\begin{align}
    \frac{1}{2}\boldsymbol{\sigma}_i^\top\boldsymbol{\widehat{Q}}\boldsymbol{\sigma}_i+\boldsymbol{\sigma}_i^\top\boldsymbol{\widehat{M}}\boldsymbol{u}_i=\frac{1}{2}\left(\boldsymbol{\sigma}_i+\boldsymbol{\widehat{Q}}^{-1}\boldsymbol{\widehat{M}}\boldsymbol{u}_i\right)^{\!\!\top}\!\boldsymbol{\widehat{Q}}\left(\boldsymbol{\widehat{Q}}^{-1}\boldsymbol{\widehat{M}}\boldsymbol{u}_i+\boldsymbol{\sigma}_i\right)-\frac{1}{2}\boldsymbol{u}_i^\top\boldsymbol{\widehat{M}}^\top\boldsymbol{\widehat{Q}}^{-1}\boldsymbol{\widehat{M}}\boldsymbol{u}_i,\nonumber
\end{align} we see that \begin{align}
    \langle\sigma_i^a\rangle &=\sum_{b=1}^n\sum_{k=1}^K(-\widehat{Q}^{-1})_{ab}\,\widehat{M}_{bk}u_i^k, & \frac{1}{N}\sum_{i=1}^N\left(\langle\sigma_i^a\sigma_i^b\rangle-\langle\sigma_i^a\rangle\langle\sigma_i^b\rangle\right)&=(-\widehat{Q}^{-1})_{ab}.
\end{align} Carrying out the Gaussian integral, \begin{align}
    z_i=e^{-\frac{1}{2}\boldsymbol{u}_i\widehat{\boldsymbol{M}}^\top\widehat{\boldsymbol{Q}}^{-1}\widehat{\boldsymbol{M}}\boldsymbol{u}_i^\top+\frac{1}{2}\ln\det(-\widehat{\boldsymbol{Q}}^{-1})+\frac{n}{2}\ln2\pi},\nonumber
\end{align}and, using $\Vert \boldsymbol{u}_k\Vert^2=N$,
\begin{align}
    \frac{1}{N}\sum_{i=1}^N\ln z_i=-\frac{1}{2}\mathrm{tr}(\widehat{\boldsymbol{M}}^\top\widehat{\boldsymbol{Q}}^{-1}\widehat{\boldsymbol{M}})+\frac{1}{2}\ln\det(-\widehat{\boldsymbol{Q}}^{-1})+\frac{n}{2}\ln2\pi.\nonumber
\end{align}The hatted variables can be eliminated,\begin{align}
    -\widehat{\boldsymbol{Q}}^{-1}\widehat{\boldsymbol{M}}&=\boldsymbol{M}, & -\widehat{\boldsymbol{Q}}^{-1}&=\boldsymbol{Q}-\boldsymbol{M}\boldsymbol{M}^\top.
\end{align}
Since $\widehat{Q}_{aa}$ is gone, we introduce $\boldsymbol{D}_\mu=\mathrm{diag}(\mu_a)_{a=1}^n$ to be chosen so that $Q_{aa}=1$, $\forall a$. In terms of $\boldsymbol{Q}$ and $\boldsymbol{M}$ we have \begin{align}
    \langle\sigma_i^a\rangle&=\sum_{k=1}^KM_{ak}u_i^k, &\frac{1}{N}\sum_{i=1}^N\left(\langle\sigma_i^a\sigma_i^b\rangle-\langle\sigma_i^a\rangle\langle\sigma_i^b\rangle\right)&=Q_{ab}-\sum_{k=1}^KM_{ak}M_{bk}.
\end{align}
The large $N$ limit of $\frac{1}{N}\ln \overline{Z(\boldsymbol{J})^n}$ is given, to leading exponential order in $N$, by the extremum in $Q_{ab}$ ($a<b$), $M_{ak}$, of \begin{align}\label{eq:log-marginal-likelihood}
    \Phi(\boldsymbol{Q},\boldsymbol{M})=\frac{1}{4\gamma T}\mathrm{tr}(\boldsymbol{X}^2)+\frac{T}{4\gamma}\mathrm{tr}(\boldsymbol{Q}^2)+\frac{1}{2\gamma}\mathrm{tr}(\boldsymbol{M}\boldsymbol{D}_\chi \boldsymbol{M}^\top)+\frac{1}{2}\mathrm{tr}\ln(\boldsymbol{Q}-\boldsymbol{M}\boldsymbol{M}^\top)-\frac{1}{2}\mathrm{tr}(\boldsymbol{D}_\mu\left(\boldsymbol{Q}-\boldsymbol{I}_n\right)),
\end{align} up to the additive constant $\frac{n}{2}\ln(2\pi e)$ which we absorbed by renormalizing $Z(\boldsymbol{J})$, $Z(\boldsymbol{0})=1$. For $n<0$, maximizing is the correct prescription. Saddle points of $\Phi$ must be locally stable to fluctuations in replica space. Beyond stability, the cumulant generating function $\Phi(n)$ must be convex in $n$, with $\Phi(0)=0$, see \cite{Kondor1983}. 

\section{B. Saddle points of the marginal likelihood}
Saddle points of $\Phi$ satisfy
\begin{equation}\label{eq:stationary-points}
   0=\begin{cases}
       \frac{\partial \Phi}{\partial \boldsymbol{Q}}\\
       \frac{\partial  \Phi}{\partial \boldsymbol{M}}
   \end{cases} \iff \begin{cases}
       (\boldsymbol{Q}-\boldsymbol{M}\boldsymbol{M}^\top)^{-1}=\boldsymbol{D}_{\mu}-\tfrac{T}{\gamma}\boldsymbol{Q}, \\
       (\boldsymbol{Q}-\boldsymbol{M}\boldsymbol{M}^\top)^{-1}\boldsymbol{M}_{k}=\frac{\chi_k}{\gamma}\boldsymbol{M}_{k}, \qquad k=1,\dots,d.
   \end{cases}
\end{equation}
The saddle-points $\boldsymbol{Q}$ and $\boldsymbol{M}$ of $\Phi$ characterize expected statistics of data configurations generated by a model randomly sampled from $P_T(\boldsymbol{J}\vert\mathcal{D})$. Using the relation $Y=\overline{Z(\boldsymbol{J})^n}$, expected statitics of model $\boldsymbol{J}$ sampled from can be computed from the knowledge of $\boldsymbol{Q}$ and $\boldsymbol{M}$. From Eq. \eqref{eq:marginal-likelihood} we indeed see that, conditionally on the $\boldsymbol{\sigma}^a$'s, finite collections of the matrix elements, $J_{ij}$'s, are Gaussian variables with conditional mean and covariance given exactly for any $N$ by \begin{align}
    \langle J_{ij}\,\vert\,\boldsymbol{\sigma}^1,\dots,\boldsymbol{\sigma}^n\rangle&=\frac{K}{N\gamma}\left(\frac{1}{K}\sum_{k=1}^K\xi_i^k\xi_j^k-\frac{1}{n}\sum_{a=1}^n\sigma_i^a\sigma_j^a\right), &i\neq j,\nonumber\\
    \langle J_{ij}J_{k\ell}\,\vert\,\boldsymbol{\sigma}^1,\dots,\boldsymbol{\sigma}^n\rangle-\langle J_{ij}\,\vert\,\boldsymbol{\sigma}^1,\dots,\boldsymbol{\sigma}^n\rangle\langle J_{k\ell}\,\vert\,\boldsymbol{\sigma}^1,\dots,\boldsymbol{\sigma}^n\rangle&=\frac{T}{N\gamma}\delta_{(ij)(k\ell)}, &i\neq j, k\neq \ell.\nonumber
\end{align}Averaging over the $\boldsymbol{\sigma}^a$'s, we get identities for the unconditional mean and covariance \begin{align}\label{eq:mean-cov-Jij}
    \langle J_{ij}\rangle&=\frac{K}{N\gamma}\left(\frac{1}{K}\sum_{k=1}^K\xi_i^k\xi_j^k-\frac{1}{n}\sum_{a=1}^n\langle\sigma_i^a\sigma_j^a\rangle\right), &i\neq j,\nonumber\\
    \langle J_{ij}J_{k\ell}\rangle-\langle J_{ij}\rangle\langle J_{k\ell}\rangle&=\frac{T}{N\gamma}\delta_{(ij)(k\ell)}+\left(\frac{T}{N\gamma}\right)^2\sum_{a,b=1}^n\left(\langle\sigma_i^a\sigma_j^a\sigma_k^b\sigma_\ell^b\rangle-\langle\sigma_i^a\sigma_j^a\rangle\langle\sigma_k^b\sigma_\ell^b\rangle\right), &i\neq j,k\neq \ell.\
\end{align}
For $N$ large, the mean and covariance of the $J_{ij}$'s, Eqs. \eqref{eq:mean-cov-Jij}, can be related to the mean and covariance of the $\sigma^a$'s, \begin{align}\label{eq:large-N-mean-cov-Jij}
    \langle J_{ij}\rangle&=\frac{K}{N\gamma}\sum_{k,\ell=1}^Ku_i^k\left(\frac{\chi_k}{K}\delta_{k\ell}-\frac{1}{n}\sum_{a=1}^nM_{ak}M_{a\ell}\right)u_j^\ell, &i\neq j,\\
    \langle J_{ij}J_{k\ell}\rangle-\langle J_{ij}\rangle\langle J_{k\ell}\rangle&=\frac{T}{N\gamma}\delta_{(ij)(k\ell)}+\left(\frac{T}{N\gamma}\right)^2\sum_{a,b=1}^n\left(\langle\sigma_i^a\sigma_j^a\sigma_k^b\sigma_\ell^b\rangle-\langle\sigma_i^a\sigma_j^a\rangle\langle\sigma_k^b\sigma_\ell^b\rangle\right), &i\neq j, k\neq \ell.
\end{align} The last term vanishes $(i,j)$ and $(k,\ell)$ are disjoint. If $(i,j)$ and $(k,\ell)$ share exactly one vertex, e.g. $i=k$, $j\neq \ell$, \begin{align}
    \langle\sigma_i^a\sigma_j^a\sigma_k^b\sigma_\ell^b\rangle-\langle\sigma_i^a\sigma_j^a\rangle\langle\sigma_k^b\sigma_\ell^b\rangle&=\left(\langle\sigma_i^a\sigma_i^b\rangle-\langle\sigma_i^a\rangle\langle\sigma_j^a\rangle\right)\langle\sigma_j^a\rangle\langle\sigma_\ell^b
    \rangle\nonumber\\
    &=\left(Q_{ab}-\sum_{r=1}^KM_{ar}M_{br}\right)\sum_{s,t=1}^Ku_j^sM_{as}M_{bt}u_\ell^t,
\end{align}and permutations thereof for the other choices $i=\ell$, $j\neq k$, and $i\neq\ell$, $j=k$, and $i\neq k, j=\ell$. If $(i,j)=(k,\ell)$, then \begin{align}
   &\langle\sigma_i^a\sigma_j^a\sigma_k^b\sigma_\ell^b\rangle-\langle\sigma_i^a\sigma_j^a\rangle\langle\sigma_k^b\sigma_\ell^b\rangle=\langle\sigma_i^a\sigma_i^b\rangle\langle\sigma_j^a\sigma_j^b\rangle-\langle\sigma_i^a\sigma_j^a\rangle\langle\sigma_i^b\sigma_j^b\rangle\nonumber\\&=\left(Q_{ab}-\sum_{m=1}^KM_{am}M_{bm}\right)\left(Q_{ab}-\sum_{n=1}^KM_{an}M_{bn}+\left(\sum_{q,r=1}^Ku_i^qM_{aq}M_{br}u_i^r+\sum_{s,t=1}^Ku_j^sM_{as}M_{bt}u_j^t\right)\right).
\end{align} 

We now study Eqs. \eqref{eq:stationary-points}. In general $\boldsymbol{Q}$ and $\boldsymbol{D}_{\mu}$ do not commute, but if we make the uniform Ansatz $\boldsymbol{D}_{\mu}=\mu \boldsymbol{I}_n$, then the saddle-point expression for $\Phi$ depends on $\boldsymbol{Q}$ and $\boldsymbol{M}$ only through rotational invariants in replica space. Upon changing variables from matrix elements to eigenvalues $\nu_a$ and eigenvectors of $\boldsymbol{Q}$, the constraints $Q_{aa}=1$ contribute \begin{align}\label{eq:I_n}
    \mathcal{I}_n(\boldsymbol{\nu})=\int_{O(n)}\mathcal{D}\boldsymbol{\Omega}\,\prod_{a=1}^n\delta\!\left(\sum_{b=1}^n\nu_b\Omega_{ab}^2-1\right),
\end{align} where $\mathcal{D}\boldsymbol{\Omega}$ denotes the flat Haar measure over the orthogonal group $O(n)$. We expect $\mathcal{I}_n(\boldsymbol{\nu})=\mathcal{O}(e^{n^2})$, which for $n=\mathcal{O}(1)$ can be neglected. Hence $\overline{Z(\boldsymbol{J})^n}$ is, to leading order in $N$, rotationally invariant in replica space. The change of variables also contributes a negligible rotationally invariant Jacobian factor $\Delta(\boldsymbol{\nu})=\mathcal{O}(e^{n^2})$ given by \begin{align}\label{eq:nu-Vandermonde}
    \Delta(\boldsymbol{\nu})=\prod_{1\le a<b\le n}\vert\nu_a-\nu_b\vert.
\end{align}

We may thus take $\boldsymbol{D}_{\mu}=\mu \boldsymbol{I}_n$. Then, by Eqs. \eqref{eq:stationary-points}, nonzero columns $\boldsymbol{M}_k$ are eigenvectors of $\boldsymbol{Q}$. Let $d\le K$ be their number, $m_k^2$ their squared norms, and $\nu_k$ their eigenvalues at a saddle point. Beware that eigenvalues of $\boldsymbol{Q}$ and $m_k^2$'s may turn out negative when $n\to-K/T$. We make the Ansatz that the restriction of $\boldsymbol{Q}$ to the orthogonal complement of the subspace spanned by the $\boldsymbol{M}_k$'s is isotropic, with saddle point eigenvalue $\nu_0$.

From \eqref{eq:stationary-points}, $\nu_k$ and $m_k^2$ ($k\le d$) obey the relations
\begin{align}\label{eq:nu-m_ell}
    \mu&=\frac{T}{\gamma}\nu_k+\frac{\chi_k}{\gamma}, &\frac{1}{\nu_k-m_k^2}&=\frac{\chi_k}{\gamma}.
\end{align}
Meanwhile, the transverse eigenvalue $\nu_0$ obeys
\begin{equation}\label{eq:nu-quadratic}
    \mu = \frac{T}{\gamma}\nu_0+\frac{1}{\nu_0}.
\end{equation}
Assuming $\nu_0$ is known, we can write saddle point expressions for the remaining quantities, such as the magnetizations
\begin{align}\label{eq:RS-FM-sol}
    \frac{m_k^2}{n}&=\frac{\chi_k}{K}\left(1-\frac{\gamma}{\chi_k\nu_0}\right)\left(1-\frac{T\nu_0}{\chi_k}\right).
\end{align}
The spherical constraint imposes a compatibility condition between the transverse and longitudinal subspaces,\begin{align}\label{eq:soft-spherical}
    n=\sum_{k\le d}\nu_k+(n-d)\nu_0,
\end{align}
from which $\nu_0$ can be determined. Enforcing this constraint leads to a quadratic equation 
whose physical root is\begin{align}
    \nu_0(d)=\frac{1}{2K}\Bigg(K-\sum_{k\le d}\chi_k+\sqrt{\Big(K-\sum_{k\le d}\chi_k\Big)^2+4dK\gamma}\Bigg),
\end{align}provided $\nu_0\le\sqrt{\gamma/T}$. We have $\nu_0(K)=\sqrt{\gamma}$ for $\sum_{k\le K}\chi_k=K$ and note that $\nu_0(d)$ is independent of $T$, see Eq. \eqref{eq:stieltjes}.

It remains to determine $d>0$. Self consistency imposes $m_d^2/n>0$. Local stability to transverse and longitudinal fluctuations requires $\nu_0(d)<\sqrt{\gamma/T}$ and $\nu_0(d)<\gamma/\chi_{d+1}$ (hence $m_{d+1}^2/n<0$) respectively, see Eqs. \eqref{eq:H-lambda-tranverse}, \eqref{eq:H-lambda-longitudinal}. Combining these requirements leads to the condition\begin{align}\label{eq:d-criterion}
    \frac{\gamma}{\chi_{d}}<\nu_0(d)<\min\left\{\frac{\gamma}{\chi_{d+1}},\sqrt{\frac{\gamma}{T}}\right\},
\end{align}which extends Eq. \eqref{eq:MAP-degeneracy} obtained for $T=0$. 

The paramagnetic solution corresponds to $d=0$,\begin{align}
    \nu_0(0)&=1, && \mu = 1+\frac{T}{\gamma}.
\end{align}
Local stability to transverse and longitudinal fluctuations requires $T<\gamma$ and $\gamma>\chi_1^2$, see Eqs. \eqref{eq:H-lambda-tranverse}, \eqref{eq:H-lambda-longitudinal}.

We now look for the marginally stable spin-glass solution \begin{align}
    \nu_0&=\sqrt{\frac{\gamma}{T}}.
\end{align}
The boundary of the marginal phase is then given by $T>T_c(\gamma,d)$ with \begin{align}\label{eq:T_c}
    T_c(\gamma,d)=\frac{\gamma}{\nu_0(d)^2}.
\end{align}When $d=0$ we get $T_c(0,\gamma)=\gamma$, consistently with the transverse stability of the $d=0$ solution $\nu_0=1$ for $T<\gamma$. It follows that $\nu_0<1$ in the marginal phase. By contrast, when $d=K$, we get $T_c(K,\gamma)=1$.

In the marginal phase, the spherical constraint, Eq. \eqref{eq:soft-spherical}, can be fulfilled only at the critical replica number $n=n_c$, \begin{align}
    n_c=-\frac{K}{T}\;\frac{\sum_{k\le d}\frac{\chi_k}{K}\left(1-\frac{\sqrt{T\gamma}}{\chi_k}\right)}{1-\sqrt{\gamma/T}}.
\end{align}This expression for $n_c$ makes sense provided the prefactor of $-\frac{K}{T}$ is positive, which happens with $T>T_c(d,\gamma)$ such that $m_k^2/n_c>0$ ($k\le d$). Since $T_c(d,\gamma)\ge\gamma$, that prefactor is $<1$ when $T>T_c(\gamma,d)$, hence $n_c>-K/T$.

We introduce the replica excess $h(\nu_0,n)=n-\text{tr}(\boldsymbol{Q})$, which at a saddle reads\begin{align}
    h(\nu_0,n)=Tn(1-\nu_0)-\left(\frac{d\gamma}{\nu_0}-\sum_{k\le d}\chi_k\right).\nonumber
\end{align} Performing implicit differentiation of $h(\nu_0,n)$ with respect to $n$ reveals that \begin{align}
    \frac{\mathrm{d}\nu_0}{\mathrm{d}n}\Big\vert_{n=n_c}=-\frac{\partial_{n}h}{\partial_{\nu_0}h}\Big\vert_{n=n_c}=-\frac{1-\nu_0}{d-n_c}<0.\nonumber
\end{align}Therefore, decreasing $n$ increases $\nu_0$, which causes transverse instabilities.

For $n<n_c\le 0$, the correct solution is obtained by freezing the marginally stable solution $\nu_0(n_c(T),T)$, defined by the vanishing of the transverse Hessian eigenvalue, see Eq. \eqref{eq:H-lambda-tranverse}, and of the replica excess, $h(\nu_0,n_c)=0$. The magnetizations are then obtained as $m_k^2(\nu_0(n_c(T),T))/n_c$ and $d$ is fixed by $d=\#\{k\le K:m_k^2/n_c>0\}$.

One can show that $n_c$ gives the slope of the rate function for the $O(N)$ large deviations of the free energy above its typical value, evaluated at the freezing point which signals the transition to $O(N^2)$ large deviations, see Sec.~D. Viewing $\Phi(n)$ as the analytic continuation of the cumulant generating function of the free energy for $n<n_c$, we expect it to be linear in $n$, see \cite{Pastore2019} for the case $n_c=0$. The resulting $\Phi(n)$ is then convex in $n$. Specifically, the prescription is\begin{align}
    \Phi(n)=\begin{cases}
        \Phi(n), & n\ge n_c,\\
        \Phi(n_c)+(n-n_c)\Phi'(n_c), & n<n_c\le 0,
    \end{cases}
\end{align}
where it is understood that these expressions are evaluated at a saddle point. Using random matrix theory, we can prove that the prescription above actually gives the correct large $N$ limit of $\frac{1}{N}\log Y$, see Sec.~D below and \cite{Jorge}.

\section{C. Eigenvalues of the Hessian}
We are interested in the local stability of stationary points of $\Phi$. For that, we study the Hessian, \begin{align}
    H&=\begin{pmatrix}
        H_{QQ} & H_{QM}\\
        H_{QM}^\top & H_{MQ}
    \end{pmatrix}, 
    & (H_{QQ})_{ab,cd} &= \frac{\partial\Phi}{\partial Q_{ab}\partial Q_{cd}}, 
    & (H_{QM})_{ab,ck} &= \frac{\partial\Phi}{\partial Q_{ab}\partial M_{ck}}, 
    & (H_{MM})_{ak,cl} &= \frac{\partial\Phi}{\partial M_{ak}\partial M_{cl}},\nonumber
\end{align}
at saddle points of $\Phi$. We denote $S_{ab}=Q_{ab}-\sum_{k}M_{ak}M_{bk}$. After some algebra, we find
\begin{align}
    (H_{QQ})_{ab,cd}&=\frac{T}{\gamma}\delta_{ac}\delta_{bd}-(S^{-1})_{ac}\;(S^{-1})_{bd},\nonumber\\
    (H_{QM})_{ab,ck}&=(S^{-1})_{ac}\sum_{j=1}^n(S^{-1})_{bj}M_{jk}+(S^{-1})_{bc}\sum_{j=1}^n(S^{-1})_{aj}M_{jk},\nonumber\\
    \frac{1}{2}(H_{MM})_{ak,bl}&=\frac{\chi_k}{\gamma}\delta_{kl}\delta_{ab}-(S^{-1})_{ab}\,\delta_{kl}-(S^{-1})_{ab}\sum_{i,j=1}^nM_{ik}(S^{-1})_{ij}M_{jl}-\sum_{i=1}^n(S^{-1})_{ai}M_{ik}\sum_{j=1}^n(S^{-1})_{bj}M_{jl}.\nonumber
\end{align}
We must diagonalize $H$ inside a space of fluctuations $(\delta Q, \delta M)$ with $\delta Q=\delta Q^T$ and $\mathrm{tr}(\delta Q)=0$, and consider\begin{align}\label{eq:delta2-F}
    \delta^2 \Phi=\sum_{a<b}\sum_{c<d}\delta Q_{ab}(H_{QQ})_{ab,cd}\delta Q_{cd}+2\sum_{a<b}\sum_{ck}\delta Q_{ab}(H_{QM})_{ab,ck}\delta M_{ck} +\sum_{ak}\sum_{bl}\delta M_{ak}(H_{MM})_{ak,bl}\delta M_{bl}.
\end{align} The starting point is to make use of the fact that rotating a saddle point of $\Phi$ in replica space gives another equally valid saddle point. We rotate to a basis where $\boldsymbol{Q}$ and $\boldsymbol{S}$ are simultaneously diagonal, and the $\boldsymbol{m}_k$ are aligned with the standard basis vectors in replica space, $M_{ak}=m_k\delta_{ak}$ for $k\le d$ and $M_{ak}=0$ for $k>d$, completed with basis vectors in the $(n\!-\!d)$-dimensional subspace orthogonal to the $\boldsymbol{m}_k$'s. In such a basis, the Hessian blocks take the form
\begin{align}
    (H_{QQ})_{ab,cd}&=\left(\frac{T}{\gamma}-\frac{1}{s_as_b}\right)\delta_{ac}\delta_{bd},\nonumber\\
    (H_{QM})_{ab,ck}&=\frac{m_k}{s_as_b}\left(\delta_{ac}\delta_{bk}+\delta_{bc}\delta_{ak}\right),\nonumber\\
    \frac{1}{2}(H_{MM})_{ak,bl}&=\left(\frac{\chi_k}{\gamma}-\frac{1}{s_a}-\frac{m_k^2}{s_as_k}\right)\delta_{ab}\delta_{kl}-\frac{m_km_l}{s_as_b}\delta_{ak}\delta_{bl}.\nonumber
\end{align}We now solve the eigenvalue problem\begin{align}
    H\begin{pmatrix}
        \delta Q\\
        \delta M
    \end{pmatrix}=\lambda\begin{pmatrix}
        \delta Q\\
        \delta M
    \end{pmatrix} &\iff\begin{cases}
        \; (H_{QQ}\delta Q)_{ab}+(H_{QM}\delta M)_{ab}&=\lambda\delta Q_{ab}, \\
        \; (H_{MQ}\delta Q)_{ck}+(H_{MM}\delta M)_{ck}&=\lambda\delta M_{ck}.
    \end{cases}\nonumber
\end{align}
For the action of $H_{QQ}$ and $H_{MQ}$ on $\delta Q$, and of $H_{MM}$ and $H_{QM}$ on $\delta M$, we find\begin{align}
    (H_{QQ}\,\delta Q)_{ab}&=\sum_{c<d}(H_{QQ})_{ab,cd}\,\delta Q_{cd}=\left(\frac{T}{\gamma}-\frac{1}{s_as_b}\right)\delta Q_{ab},\nonumber\\
    (H_{QM}\delta M)_{ab}&=\sum_{ck}(H_{QM})_{ab,ck}\,\delta M_{ck}=\frac{m_b}{s_as_b}\delta M_{ab}+\frac{m_a}{s_as_b}\delta M_{ba},\nonumber\\
    (H_{MQ}\delta Q)_{ck}&=\sum_{a<b}(H_{QM})_{ab,ck}\,\delta Q_{ab}=\frac{2m_k}{s_cs_k}\delta Q_{ck},\nonumber\\
    \frac{1}{2}(H_{MM}\delta M)_{ak}&=\frac{1}{2}\sum_{bl}(H_{MM})_{ak,bl}\,\delta M_{bl}=\left(\frac{\chi_k}{\gamma}-\frac{1}{s_a}-\frac{m_k^2}{s_as_k}\right)\delta M_{ak}+\frac{m_k}{s_a}\delta_{ak}\sum_{l}\frac{m_l}{s_l}\delta M_{ll}.\nonumber
\end{align}
We see that $H$ has $\frac{1}{2}(n-d)(n-d-1)$ purely transverse modes, $(\delta Q,\delta M)$ with $\delta Q_{ab}>0$ for $a,b>d$, $a\neq b$, and $\delta Q_{ab}=0$ otherwise (so that $(H_{QM}\delta Q)_{ab}=0$), and $\delta M_{ck}=0$ (hence $(H_{MM}\delta M)_{ak}=0$). The corresponding eigenvalues $\lambda_{ab}$ can be read off directly from $(H_{QQ}\delta Q)_{ab}$ above\begin{align}\label{eq:H-lambda-tranverse}
    \lambda_{ab}=\frac{T}{\gamma}-\frac{1}{s_as_b}, \qquad a,b>d, \; a\neq b. 
\end{align}
It is also clear that there are $n(K-d)$ modes with $\delta M_{ak}>0$ for any $a$ and $k>d$, $\delta M_{ak}=0$ for any $a$, $k\le d$, and $\delta Q_{ab}=0$ for all $a,b$, with eigenvalues $\lambda_{ak}$, where \begin{align}\label{eq:H-lambda-longitudinal}
    \lambda_{ak}=2\left(\frac{\chi_k}{\gamma}-\frac{1}{s_a}\right), \qquad k>d.
\end{align}
$H$ has $d$ other modes with $\delta M_{kk}>0$ for $k\le d$ and $\delta M_{ak}=0$ otherwise, and $\delta Q_{ab}=0$. The associated eigenvalues are eigenvalues of the rank-one deformation $H_{d}$ in $d$ dimensions of a diagonal matrix with non-uniform diagonal, \begin{align}
H_{d,kl}=2\left(\frac{\chi_k}{\gamma}-\frac{1}{s_k}-\frac{m_k^2}{s_k^2}\right)\delta_{kl}-2\frac{m_km_l}{s_ks_l}. \nonumber
\end{align} The eigenvalues are the real roots of \begin{align}
    1-\sum_{k=1}^d\frac{m_k^2/s_k^2}{\chi_k/\gamma-1/s_k-m_k^2/s_k^2-\lambda/2}=0.
\end{align}
We also find $2(n-d)d$ modes with $\delta Q_{ak},\delta M_{ak}>0$ for $k\le d$ and $a>d$, and $\delta Q_{ab}=\delta M_{ak}=0$ otherwise. Indeed $\delta Q_{ak}$ talks only to $\delta M_{ak}$ since $m_a=0$, and $\delta M_{ak}$ talks only back to $\delta Q_{ak}$. On the subspace spanned by these vectors, $H$ acts as the $2\times 2$ matrix $H_{2}$, with \begin{align}
    H_{2}&=\begin{pmatrix}
       \frac{T}{\gamma}-\frac{1}{s_as_k} & \frac{m_k}{s_as_k} \\
       \frac{m_k}{s_as_k} & 2\left(\frac{\chi_k}{\gamma}-\frac{1}{s_a}-\frac{m_k^2}{s_as_k}\right)
    \end{pmatrix}.\nonumber
\end{align}The eigenvalues $\lambda_{ak,\pm}$ are given by \begin{align}
    \!\lambda_{ak,\pm}&=\frac{1}{2}\left(\frac{T}{\gamma}\!-\!\frac{1}{s_as_k}\!+\!2\left(\frac{\chi_k}{\gamma}\!-\!\frac{1}{s_a}\!-\!\frac{m_k^2}{s_as_k}\right)\!\pm\!\sqrt{\left(\frac{T}{\gamma}\!-\!\frac{1}{s_as_k}\!-\!2\left(\frac{\chi_k}{\gamma}\!-\!\frac{1}{s_a}\!-\!\frac{m_k^2}{s_as_k}\right)\right)^2\!+\!\frac{4m_k^2}{s_a^2s_k^2}}\right),\quad a>d, k\le d.
\end{align}
Lastly, $H$ has $\frac{1}{2}d(d-1)$ modes $(\delta Q,\delta M)$ with $\delta Q_{kl}, \delta M_{kl}, \delta M_{lk}>0$ and zero for the remaining components. The eigenvalues are the $\frac{3}{2}d(d-1)$ eigenvalues of the $3\times 3$ matrices $H_{3,kl}$, \begin{align}
    H_{3,kl}=\begin{pmatrix}
        \frac{T}{\gamma}-\frac{1}{s_ks_l} & \frac{m_l}{s_ks_l} & \frac{m_k}{s_ks_l}\\
        \frac{m_l}{s_ks_l} & 2\left(\frac{\chi_l}{\gamma}-\frac{1}{s_k}-\frac{m_l^2}{s_ks_l}\right) & 0 \\
        \frac{m_k}{s_ks_l} & 0 & 2\left(\frac{\chi_k}{\gamma}-\frac{1}{s_l}-\frac{m_k^2}{s_ks_l}\right)
    \end{pmatrix}.
\end{align}
Summing the multiplicities gives $\frac{1}{2}n(n-1)+nK$, the size of $H$, as it should.

\section{D. Large deviations of the ground state energy and of the free energy}

\subsection{Large deviations of the largest eigenvalue of a rank-one deformed Wigner matrix}
Let $\boldsymbol{u}$ be a deterministic unit vector of size $N$. Over the course of the present section, we assume that $\boldsymbol{J}$ is drawn from the rank-one deformed Gaussian orthogonal ensemble, \begin{align}\label{eq:rank-one-deformed-GOE}
    \boldsymbol{J}&=\boldsymbol{W}+\theta\,\boldsymbol{u}\boldsymbol{u}^\top,
\end{align} where $\boldsymbol{W}$ is a Wigner matrix of parameter $\sigma^2$, meaning $W_{ij}\sim\mathcal{N}(0,\frac{\sigma^2}{N}(1+\delta_{ij}))$ with $W_{ji}=W_{ij}$. The limiting spectral density of $\boldsymbol{W}$ is a semicircle, of radius $2\sigma$ with these conventions, centered around zero. 

Matrices of the form \eqref{eq:rank-one-deformed-GOE} are distributed with density $\rho_J(\boldsymbol{J})$, Eq. \eqref{eq:marginal-likelihood} for any $K$ as long as there is $D=1$ nonzero eigenvalue $\chi_1$, with $\theta=1/\gamma$, $\sigma^2=T/\gamma$.

The typical value $\lambda_{\text{typ}}(\sigma,\theta)$ of $\lambda_{\max}$ satisfies \cite{Kosterlitz1976, Edwards1976} \begin{align}
    \lambda_{\text{typ}}(\sigma,\theta)=\begin{cases}
      2\sigma, &  \theta < \sigma,\\
      \theta+\frac{\sigma^2}{\theta}, &  \theta > \sigma.
    \end{cases}\nonumber
\end{align}  
We look for a large deviation principle (LDP) with rate function $\Omega$ of speed $N$ for the largest eigenvalue $\lambda_{\max}$ of $\boldsymbol{J}$, \begin{align}
    \mathbb{P}(\lambda_{\max}\approx \lambda)&\asymp e^{-N\Omega(\lambda)}, &\lambda&\ge 2\sigma.\nonumber
\end{align}

Large deviations of extreme eigenvalue of random matrices are asymmetric \cite{Dean2006, Majumdar2009}. Large deviations to the right of the bulk edge can be achieved by pulling a single eigenvalue out of the bulk over a finite distance. This is the event of drawing a random matrix with an unually large projection along a single direction, whose probability decays exponentially with $N$ \cite{Majumdar2009}. Meanwhile, large deviations to the left require to push a finite fraction of the eigenvalues over a finite distance over the support of the bulk. This is the event of drawing a random matrix with an unusually small projection along a finite fraction of directions, whose probability is suppressed exponentially with $N^2$ \cite{Dean2006}.

The speed $N$ rate function $\Omega$ be determined from the cumulant generating function, $\phi(t)$, by tilting the law of $\lambda_{\max}$, \begin{align}
    \phi(t)&\equiv \lim_{N\to+\infty}\frac{1}{N}\ln\mathbb{E}[e^{Nt\lambda_{\max}}]=\sup_{\lambda\ge2\sigma}\left\{t\lambda-\Omega(\lambda)\right\}.
\end{align}
Provided we know $\phi(t)$, the above Legendre-Fenchel transform can be inverted to recover $\Omega(\lambda)$. The rate function $\Omega(\lambda)$ is known \cite{Maida2019}. Here, we propose an independent derivation with a replica approach \cite{Fyodorov2013, Steinberg2024}. 

The starting point is the introduction of a large parameter $1/T'$, noting that \begin{align}
    \overline{Z(\boldsymbol{J}/T')^n}\simeq\int_{2\sigma}^{\infty}\!\mathrm{d}\lambda\; e^{N\left(\frac{n}{2T'}\lambda-\Omega(\lambda)\right)}.\nonumber
\end{align} We see that we must compute the averaged replicated partition function for small $n$ with $t=n/2T'$ fixed as $T'\to0^+$.
Under a Replica Symmetric Ansatz, \begin{align}
    \lim_{N\to+\infty}\frac{1}{N}\ln\overline{Z(\boldsymbol{J}/T')^n}&=\frac{1}{2}\inf_{q,m}V(q,m,n), \nonumber\\
    V(q,m,n)&=\frac{n\sigma^2}{2T'^2}(1+(n-1)q^2)+\frac{n}{T'}\theta m^2+(n-1)\ln(1-q)+\ln(1-q+n(q-m^2)).\nonumber
\end{align} Expanding $V$ to small $n$ for $T'$ fixed gives $V(q,m,n)=n V_1(q,m)$ with $V_1$ such that $\partial_qV_1\vert_{q=q_0}=\partial_{m^2}V_1\vert_{m^2=m^2_0}=0$, \begin{align}
    &\begin{cases}
        q_0=1-\frac{T'}{\theta}, \\
        m_0^2=q_0\left(1-\frac{\sigma^2}{\theta^2}\right),
    \end{cases} &&\mathrm{or} &&\begin{cases}
        q_0=1-\frac{T'}{\sigma}, \\
        m_0^2=0.
    \end{cases} \nonumber
\end{align} When $T'\to0$, replicas align along the top eigenvector $\boldsymbol{v}$ of $\boldsymbol{J}$, and $m_0^2$ measures the overlap of $\boldsymbol{v}$ with the spike $\boldsymbol{u}$.

From the above, for $T'\to0$ with $n=2tT'$, we expect the solution to be of the form\begin{align}
    &\begin{cases}
        q(t)=1-T'\Delta(t),\\
        m(t)^2=\mathcal{O}(1).
    \end{cases}\nonumber
\end{align}Plugging this Ansatz, and using $\phi(t)=\frac{1}{2}V(q=1-T'\Delta,m^2=\mathcal{O}(1),n=2tT')+o(T')$, we get\begin{align}
    \phi(t)=\sigma^2t^2+\sigma^2t\Delta+\theta tm^2+\frac{1}{2}\ln\left(1+\frac{2t(1-m^2)}{\Delta}\right),\nonumber
\end{align}to be made stationary in $\Delta$ and $m^2$. 

\textbf{Branch $\boldsymbol{m^2=0}$.} For $\theta<\sigma$ there is a solution \begin{align}
    0&=\sigma^2\Delta^2+2\sigma^2 t\Delta-1, && \Delta(t)=\frac{1}{\sigma}\left(\sqrt{1+\sigma^2t^2}-\sigma t\right).
\end{align}Note $\Delta(t)\ge 0$ for all $t$. For $\sigma>\theta$ the typical value of $q$ is $q(0)=1-T'/\sigma$. Substituting $\Delta(t)$ into $\phi(t)$, \begin{align}
    \phi(t)&=\sigma t\sqrt{1+\sigma^2t^2}+\frac{1}{2}\ln\left(1+2\sigma^2t^2+2\sigma t\sqrt{1+\sigma^2 t^2}\right), & \lambda(t)&=\phi'(t)=2\sigma\sqrt{1+\sigma^2t^2}.\nonumber
\end{align}
Inverting the relation between $t$ and $\lambda$, we recover the result of \cite{Majumdar_2009} for the $\theta=0$ case, \begin{align}\label{eq:Omega0}
   t(\lambda)&=\frac{1}{2\sigma^2}\sqrt{\lambda^2-4\sigma^2}, & \Omega_0(\lambda)&=t(\lambda)\lambda-\phi(t(\lambda))=\frac{\lambda}{4\sigma^2}\sqrt{\lambda^2-4\sigma^2}-\frac{1}{2}\ln\left(\frac{\lambda}{2\sigma^2}\left(\lambda+\sqrt{\lambda^2-4\sigma^2}\right)-1\right).
\end{align}
We can now comment on this result. For $\sigma>\theta$ the typical value of $\lambda$ is $\lambda(0)=2\sigma$ with $\Omega_0(2\sigma)=0$. The optimal strategy to realise large deviations $\lambda>2\sigma$ is to generate an apparent outlier that is uncorrelated with the spike.

\textbf{Branches $\boldsymbol{{m^2>0}}$.} There are solutions
\begin{align}
    m(t)^2&=1-\frac{\sigma^2}{\theta}\Delta(t), &\Delta(t)&=\frac{1}{\theta+2\sigma^2t}, & t>t_c, &&t_c=\frac{\theta}{2\sigma^2}\left(\frac{\sigma^2}{\theta^2}-1\right)\begin{cases}
    >0 &\theta < \sigma,\\
    <0 &\theta > \sigma,
    \end{cases}\nonumber
\end{align}
where $t_c$ is defined as the least $t$ such that $m(t)^2>0$. Substituting $m(t)^2$ and $\Delta(t)$ into $\phi(t)$,
\begin{align}
    \phi(t)&=\frac{\theta^2}{4\sigma^2}\left(\rho(t)^2-1\right)+\frac{1}{2}\ln\rho(t), &\lambda(t)&=\phi'(t)=\theta\rho(t)+\frac{\sigma^2}{\theta \rho(t)},  &\rho(t)&\equiv\frac{1}{\theta\Delta(t)}=1+\frac{2\sigma^2 t}{\theta}, && t>-\frac{\theta}{2\sigma^2}.\nonumber
\end{align}The map $\rho\mapsto\theta\rho+\sigma^2/\theta\rho$ attains its minimum $2\sigma$ in $\rho_{\min}\equiv\rho(t_{\min})=\sigma/\theta$, where \begin{align}\label{eq:tmin}
    t_{\min}=\frac{\theta}{2\sigma^2}\left(\frac{\sigma}{\theta}-1\right)\begin{cases}
    >0 & \theta < \sigma,\\
    <0 & \theta > \sigma.
    \end{cases}\nonumber
\end{align} The branch $m(t)^2>0$ exists for $\rho>\rho_c\equiv\rho(t_c)=\sigma^2/\theta^2$ where $\lambda(t_c)=\theta+\sigma^2/\theta$. Thus, $\rho=\rho_{\min}$ might not be feasible.
\begin{itemize}
    \item If $\theta < \sigma$, then $\rho_c>\rho_{\min}>1$, and the range of admissible values $\rho>\rho_c$ does not include $\rho_{\min}$, resulting in \begin{align}
       &\lambda\ge \theta+\frac{\sigma^2}{\theta}, &&t\ge t_c>0.
    \end{align}
    The typical values of $\lambda$ and $m^2$ are, respectively, $2\sigma$ and $0$. The optimal strategy to realise large deviations $\lambda>\lambda(t_c)=\theta+\sigma^2/\theta$ is to make $m(t)^2>m(t_c)^2=0$. This is achieved by choosing $t$ positive, larger than $t_c$, unbounded from above, with $1-m(t)^2=\mathcal{O}(t^{-1})$ in the large $t$ limit.
    \item If $\theta>\sigma$, then $\rho_c<\rho_{\min}<1$, so the range of admissible values $\rho>\rho_c$ does include $\rho_{\min}=\sigma/\theta$. Hence \begin{align}
        &2\sigma \le \lambda \le \theta+\frac{\sigma^2}{\theta}, && t_{\min}\le t\le 0.
    \end{align}
    The typical values of $\lambda$ and $m^2$ are, respectively, $\theta+\sigma^2/\theta$ and $1-\sigma^2/\theta^2$. The optimal strategy to realise large deviations $\lambda<\lambda(0)=\theta+\sigma^2/\theta$ is to make $m(t)^2<m_0^2=1-\sigma^2/\theta^2$. This is achieved by choosing $t$ negative, allowed down to $t_{\min}$, where  $\lambda(t_{\min})=2\sigma$ with $m(t_{\min})^2=1-\sigma/\theta>0,<m_0^2$.
\end{itemize}

We have\begin{align}
    0&=\theta\rho^2-\lambda\rho+\frac{\sigma^2}{\theta}, &&\rho_\pm(\lambda)=\frac{\lambda\pm\sqrt{\lambda^2-4\sigma^2}}{2\theta}, &&\lambda\ge2\sigma.\nonumber
\end{align} For $\theta<\sigma$, we need $\rho>\rho_c$, which picks the $+$ root. For $\theta>\sigma$, we need $\rho(\lambda=\theta+\sigma^2/\theta)=1$, which also picks $+$. We thus pick $\rho(\lambda)=\rho_+(\lambda)$ regardless, with $t(\lambda)=\frac{\theta}{2\sigma^2}\left(\rho(\lambda)-1\right)$, to obtain \begin{align}\label{eq:Omega1}
    \Omega_{1}(\lambda)=t(\lambda)\lambda-\phi(t(\lambda))&=\frac{\theta^2}{4\sigma^2}\left(\rho(\lambda)-1\right)^2+\frac{\rho(\lambda)-1}{2\rho(\lambda)}-\frac{1}{2}\ln\rho(\lambda)\nonumber\\
    &=\frac{\lambda}{8\sigma^2}\left(\lambda+\sqrt{\lambda^2-4\sigma^2}\right)-\frac{1}{2}\ln\left(\frac{\lambda+\sqrt{\lambda^2-4\sigma^2}}{2\theta}\right)+\frac{1}{2\sigma^2}\left(\frac{\theta^2}{2}-\theta\lambda+\frac{\sigma^2}{2}\right).
\end{align}We see that $\Omega_1(\lambda)$ satisfies  \begin{align}
    \Omega_{1}\left(\theta+\frac{\sigma^2}{\theta}\right)=\begin{cases}
        0,&\theta>\sigma,\\
        \Omega_0\left(\theta+\frac{\sigma^2}{\theta}\right)=\frac{1}{4}\left(\frac{\sigma^2}{\theta^2}-\frac{\theta^2}{\sigma^2}\right)+\ln\left(\frac{\theta}{\sigma}\right)&\theta<\sigma.
    \end{cases}
\end{align}The global rate function $\Omega(\lambda)$ can then be reconstructed as \begin{align}
    \Omega(\lambda)=\begin{cases}
        \Omega_{1}(\lambda), & \theta > \sigma,\\
        \begin{cases}
            \Omega_0(\lambda), & 2\sigma\le \lambda<\theta+\frac{\sigma^2}{\theta},\\
            \Omega_{1}(\lambda), & \lambda \ge  \theta+\frac{\sigma^2}{\theta},
        \end{cases}& \theta < \sigma.
    \end{cases}
\end{align}

\begin{figure}[h]
    \centering
    \includegraphics[width=0.5\linewidth]{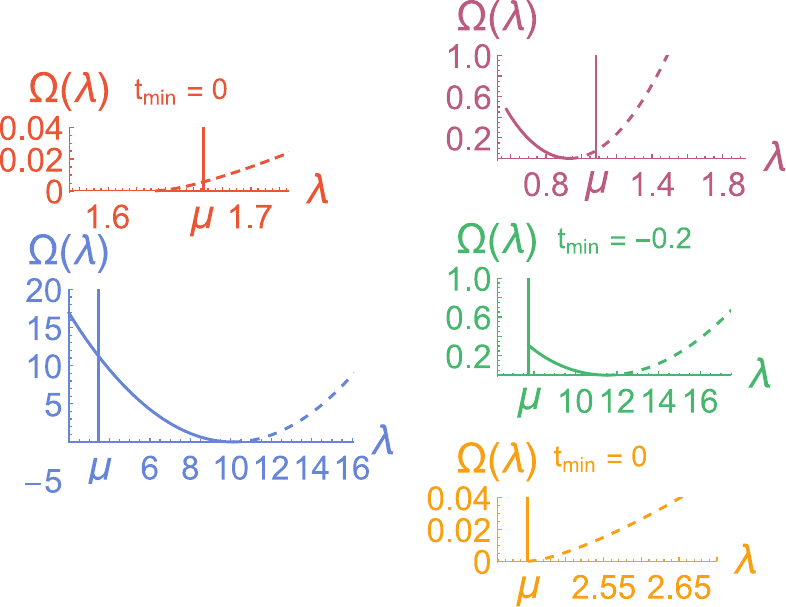}
    \caption{Rate function $\Omega$ for the speed $N$ large deviations of the largest eigenvalue $\lambda\ge\lambda_{\text{edge}}=2\sigma$ of a rank-one deformed Wigner matrix to the right (dashed) and to the left (solid) of the typical value of $\lambda_{\max}$ ($\Omega(\lambda_{\max})=0$), with the position of the chemical potential ($\mu$) in the 5 different phases. \textit{Blue}, \textit{green}, and \textit{orange} correspond to points in the $(\gamma,T)$ plane marked with $\times$ in the phase diagram of Fig.~1, Main, and $t_{\min}$ defined in Eq. \eqref{eq:tmin}. \textit{Red phase}: $\mu>\lambda_{\max}=\lambda_{\text{edge}}$ ($\gamma=1.5$, $T=1.0$). \textit{Purple phase}: $\mu>\lambda_{\max}>\lambda_{\text{edge}}$ ($\gamma=1.2$, $T=0.1$). \textit{Blue phase}: $\mu=\lambda_{\max}>\lambda_{\text{edge}}$ ($\gamma=0.1$, $T=0.1$). \textit{Green phase}: $\mu=\lambda_{\max}=\lambda_{\text{edge}}$ with $m^2>0$ ($\gamma=0.1$, $T=1.5$). \textit{Orange phase:} $\mu=\lambda_{\max}=\lambda_{\text{edge}}$ with $m^2=0$ ($\gamma=1.0$, $T=1.5$).}
    \label{fig:placeholder}
\end{figure}

\subsection{Large deviations of the rank-one deformed spherical model free energy density}
We have computed the large deviations of $\lambda_{\max}$, which is half the ground state free energy. Setting $T'=1$ and keeping $n=\mathcal{O}(1)$, we are now interested in the cumulant generating function
\begin{align}
\Phi(n)&\equiv\lim_{N\to+\infty}\frac{1}{N}\ln\mathbb{E}[e^{Nn\psi(\lambda_{\max})}]
\end{align}
of the free energy density $Ng(\boldsymbol{J})=\ln Z(\boldsymbol{J})$, given for $N$ large by the expressions
\begin{align}
    \psi(\lambda_{\max})&=\inf_{\mu> \lambda_{\max}}g(\mu), &&\text{where} &g(\mu)&=\frac{1}{2}\mu-\frac{1}{2N}\sum_{j=1}^N\ln(\mu-\lambda_j),
\end{align}with $\mu$ the Lagrange multiplier imposing the spherical constraint, the condition $\mu>\lambda_{\max}$ being enforced by an $o(N)$ contribution. We see that $\psi_0$ depends on the empirical density $\hat\rho_N=\frac{1}{N}\sum_{i=1}^N\delta_{\lambda_i}$ of all the $\lambda_j$s. Informally speaking, for samples $\boldsymbol{J}$ with a probability decaying exponentially with $N$, we expect that $\hat{\rho}_N$ remains close to its typical value, \begin{align}
    \rho_\lambda(\lambda)&=\frac{\sqrt{4\sigma^2-\lambda^2}}{2\pi\sigma^2}.
\end{align}
Through $\mu_0$, which satisfies the condition \begin{align}
    1&=\frac{1}{N}\sum_{j=1}^N\frac{1}{\mu_0-\lambda_j}, &\mu_0&>\lambda_{\max},
\end{align} the free energy density $\psi$ is a deterministic function of the random variable $\lambda_{\max}$. Large deviations, i.e., improbable fluctuations, of $\psi$ arise from the most probable of the improbable fluctuations of $\lambda_{\max}$. Contracting the LDP of $\lambda_{\max}$, \begin{align}
\mathbb{P}(\psi(\lambda)\approx \psi)&\asymp e^{-I(\psi)}, & I(\psi)&=\min_{\lambda\ge2\sigma:\psi(\lambda)=\psi}\Omega(\lambda).
\end{align}
We thus have
\begin{align}
\Phi(n)&=\sup_{\psi}\left\{n\psi-I(\psi) \right\}=\sup_{\lambda\ge2\sigma}\left\{n\psi(\mu_0(\lambda))-\Omega(\lambda)\right\},
\end{align}
where\begin{align}
    \mu_0(\lambda)&=\begin{cases}
        1+\sigma^2, & q(\lambda)\le 0,\\
        \lambda, & q(\lambda)>0,
    \end{cases}&q(\lambda)&=1-\frac{1}{2\sigma^2}\left(\lambda-\sqrt{\lambda^2-4\sigma^2}\right).\nonumber
\end{align}We note that for $\lambda_{\max}=\lambda_{\text{typ}}(\sigma,\theta)$, one has \begin{align}
    \mu_0(\sigma,\theta)=\begin{cases}
    \begin{cases}
       1+\sigma^2, & \sigma<1,\\
        2\sigma, & \sigma>1,
    \end{cases} 
        &\theta < \sigma,\\
    \begin{cases}
        1+\sigma^2, & \theta < 1,\\
        \theta+\frac{\sigma^2}{\theta},  &\theta>1,
    \end{cases}
        &\theta > \sigma.
    \end{cases}\nonumber
\end{align}
We explicitly carried out the Legendre transform to verify that the resulting expression of $\Phi(n)$ derived here agreed with its expression derived in Sec.~A. We also see that $n_c$ is the slope of $I$ near the edge. Optimizing over $\lambda$ gives\begin{align}
    n=\frac{t(\lambda)}{\psi_0'(\lambda)},
\end{align}
where we used $t(\lambda)=\Omega'(\lambda)$. We start with $\theta>\sigma$, assuming $\sigma>1$ so that $q(2\sigma)>0$. As we make $n$ more negative, the optimal value of $\lambda$ decreases, until it reaches $2\sigma$. The replica number $n$ then saturates at the critical value $n_c$,\begin{align}\label{eq:n_eff}
    n_c=\lim_{\lambda\downarrow2\sigma}\frac{t(\lambda)}{\psi'(\lambda)}=\frac{t_{\min}}{\psi'(2\sigma^+)}=\frac{\sigma-\theta}{\sigma(\sigma-1)},
\end{align}below which the value of $\psi$ cannot be decreased further with samples whose probability decays exponentially with $N$.

\begin{figure}
    \centering
    \includegraphics[width=0.5\linewidth]{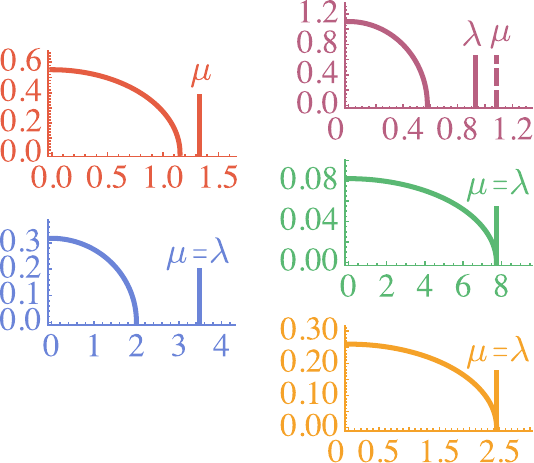}
    \caption{Theoretical spectrum of a random matrix drawn from $P_T(\boldsymbol{J}\vert\mathcal{D})$ for data lying along a one-dimensional manifold, a tilted ensemble equivalent to the large deviations of rank-one deformed Gaussian random matrices given by Eq. \eqref{eq:rank-one-deformed-GOE}, with largest eigenvalue $\lambda$ (only depicted when lying out of the bulk) and chemical potential $\mu$, in the 5 different phases of of Fig.~1 in the Main. \textit{Blue}, \textit{green}, and \textit{orange} correspond to points in the $(\gamma,T)$ plane marked in the phase diagram of Fig.~1 in the Main. \textit{Purple} corresponds to $\gamma=1.2$, $T=0.1$ as in Fig.~1 here in the SM and \textit{red} corresponds to $\gamma=3$, $T=1$.}
    \label{fig:placeholder}
\end{figure}

\section{E. Learning metrics}
\subsection{Likelihood of training data}
The likelihood of the training data is accessed by differentiating $\ln Y$ with respect to $T^{-1}$ fixing $\sigma^2=T/\gamma$, 
\begin{align}
    \mathcal{L}_{\mathrm{train}}(\boldsymbol{C})=\left\langle\frac{1}{2N}\mathrm{tr}(\boldsymbol{J}\boldsymbol{C})-\frac{1}{N}\ln Z(\boldsymbol{J})\right\rangle_{\boldsymbol{J}\vert \mathcal{D}}=\frac{1}{NK}\left(\frac{\partial\ln Y}{\partial T^{-1}}\right)_{\!\sigma^2}.
\end{align}Using the expression we found with replicas for the large $N$ limit of $\frac{1}{N}\ln Y$ we get, after simplifications, \begin{align}
    \mathcal{L}_{\mathrm{train}}(\boldsymbol{C})=
        \frac{1}{K}\Big(\frac{\partial\Phi}{\partial T^{-1}}\Big)_{\!\sigma^2}=\frac{1}{2K\gamma}\sum_{k\le K}\chi_k^2+\frac{T}{2K\gamma}\sum_{k\le d}\chi_km_k^2-\frac{T}{4\gamma}\nu_0^2-\frac{1}{2}\ln \nu_0-\frac{\mu}{2}(1-\nu_0)-\frac{1}{2},
\end{align}where $\{m_k\}_{k\le d}$ and $\{\mu,\nu_0,d\}$ are given by their appropriate respective saddle point expression, and it is understood that we are neglecting the additive term $\frac{n}{2}\ln2\pi$. In particular, \begin{align}\label{eq:Ltrain}
    \mathcal{L}_{\text{train}}=\begin{cases}
        \frac{1}{2K\gamma}\sum_{k\le K}\chi_k^2-\frac{T}{4\gamma}-\frac{1}{2} &d=0, T<\gamma, \\
        \frac{1}{2K\gamma}\sum_{k\le K}\chi_k^2-\frac{1}{4}-\frac{1}{4}\ln\frac{T}{\gamma}-\sqrt{\frac{T}{\gamma}}
        , &d=0, T>\gamma,\\
        \frac{1}{4}-\frac{T}{2}-\frac{1}{4}\ln\frac{\gamma}{T}, &d=K, T>1,\\
        -\frac{T}{4}-\frac{1}{4}\ln\gamma, &d=K, T\le 1.
    \end{cases}
\end{align}

\subsection{Likelihood of generated data}
We compute the likelihood of data $\boldsymbol{\sigma}\sim P_{\sigma\vert J}(\boldsymbol{\sigma}\vert \boldsymbol{J})$ generated by a model $\boldsymbol{J}\sim P_{T}(\boldsymbol{J}\vert \mathcal{D})$, defined as \begin{align}\label{eq:Lgen}
    \mathcal{L}_{\text{gen}}=\frac{1}{N}\langle\langle\ln P_{\sigma\vert J}(\boldsymbol{\sigma})\rangle_{\sigma\vert J}\rangle_{J\vert\mathcal{D}}=\left\langle\frac{1}{2N}\mathrm{tr}(\boldsymbol{J}\boldsymbol{\sigma}\boldsymbol{\sigma}^\top)-\frac{1}{N}\ln Z(\boldsymbol{J})\right\rangle_{J\vert \mathcal{D}}.
\end{align}
Using $\langle\boldsymbol{\sigma}\boldsymbol{\sigma}^\top\rangle_{\sigma\vert J}=(\mu \boldsymbol{I}-\boldsymbol{J})^{-1}$, we get the identity \begin{align}
    \frac{1}{N}\langle\mathrm{tr}(\boldsymbol{J}\boldsymbol{\sigma}\boldsymbol{\sigma}^\top)\rangle_{\boldsymbol{\sigma}\vert\boldsymbol{J}}=\mu - 1.\nonumber
\end{align}This follows by writing $\boldsymbol{J}=-(\mu\boldsymbol{I}-\boldsymbol{J})+\mu\boldsymbol{I}$ and remembering that $\mu$ satisfies $\mathrm{tr}(\mu \boldsymbol{I}-\boldsymbol{J})^{-1}=N$. We thus reach\begin{align}
    \mathcal{L}_{\text{gen}}=\frac{1}{2}\left(\frac{\mu^2-\mu\sqrt{\mu^2-4\sigma^2}}{4\sigma^2}+\ln\left(\frac{\mu+\sqrt{\mu^2-4\sigma^2}}{2}\right)-\frac{3}{2}\right)
\end{align} upon completing the integral over the semicircle of radius $2\sigma$ ($\sigma^2=T/\gamma$). We have that \begin{align}
    \mu=\begin{cases}
        1+\sigma^2, & d=0,\\
        2\sigma,    & d>0, T>T_c(d,\gamma),\\
        \sigma^2\nu_0(d)+\frac{1}{\nu_0(d)}, &d>0, T\le T_c(d,\gamma),
    \end{cases}
\end{align} with in particular $\mu=\frac{1+T}{\sqrt{\gamma}}$ for $d=K$, $T\le 1$. Then\begin{align}
    \mathcal{L}_{\text{gen}}=\begin{cases}
        \frac{T}{4\gamma}-\frac{1}{2}, & d=0,\\
        \frac{1}{4}\ln\frac{T}{\gamma}-\frac{1}{4}, & d>0, T>T_c(d,\gamma),\\
        \frac{T}{4}-\frac{1}{4}\ln \gamma-\frac{1}{2}, & d=K, T\le 1.
    \end{cases}
\end{align}For $d=K=1$ we notice the identity \begin{align}
    \mathcal{L}_{\text{gen}}\vert_{T=1}&=\mathcal{L}_{\text{train}}\vert_{T=1}.
\end{align}

\begin{figure}[h]
    \centering
    \includegraphics[width=0.5\linewidth]{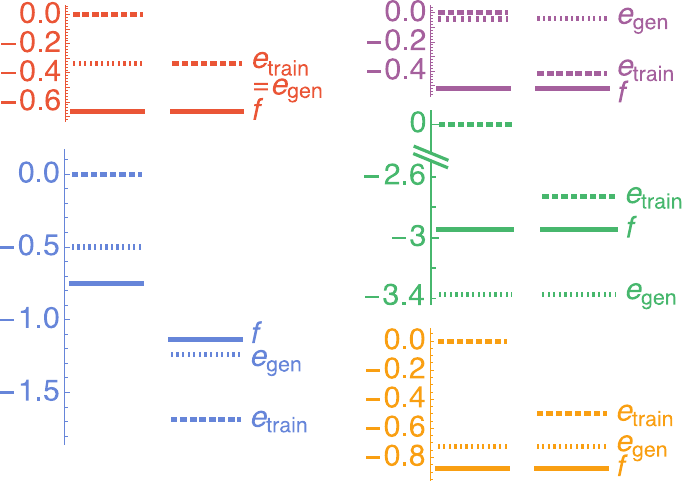}
    \caption{Average energy levels of the training data $e_\text{train}$, the generated data $e_\text{gen}$, and model free energy $f$, before (\textit{left}) and after (\textit{right}) training for the 5 phases of ensemble learning in the $(\gamma,T)$ plane, corresponding to Fig.~1 from the SM. \textit{Blue}, \textit{green}, and \textit{orange} correspond to points in the $(\gamma,T)$ plane marked with $\times$ in the phase diagram of Fig.~1 from the Main. Quantities before training are defined by averaging over $\boldsymbol{J}$ drawn from the prior $P_G(\boldsymbol{J})$. In the \textit{blue} phase, $n<0$ results in a decrease of $f$. In the \textit{red}, \textit{purple}, and \textit{orange} phases, where $m^2=0$, a decrease in $e_\text{train}$ is nonetheless observed due to a nonzero alignment of $\langle J_{ij}\rangle$ with $C_{ij}$ ($i<j$).}
    \label{fig:placeholder}
\end{figure}

\subsection{Case of nearly finite dimensional data}
We consider training data $\boldsymbol{\xi}^\ell$ perfectly distributed along a line with varying amplitudes $\Vert\boldsymbol{\xi}^\ell\Vert^2$, in the sense that $\boldsymbol{\xi}^\ell=a_\ell \boldsymbol{u}$ ($\ell=1,\dots,K$) for some direction $\boldsymbol{u}$. Then $\chi_1=\frac1N\sum_{\ell=1}^K\Vert\boldsymbol{\xi}^\ell\Vert^2$ while $\chi_{\ell>1}=0$. Replica theory predicts $d=1$ for $\gamma<\chi_1$ and $T<\gamma/\nu_0(1)^2$ or $\gamma/\nu_0(1)^2<T<\chi_1^2/\gamma$. Else, $d=0$. Because $\nu_0(1)=\frac{1}{2K}(K-\chi_1+\sqrt{(K-\chi_1)^2+4K\gamma})$, one has that $T_c(\gamma,1)=\gamma/\nu_0(1)^2$ is independent of $\gamma$ only for $\chi_1=K\iff \sum_{\ell=1}^K\Vert\boldsymbol{\xi}^\ell\Vert^2=N$. For $\frac{1}{K}\sum_{\ell=1}^K\Vert\boldsymbol{\xi}^\ell\Vert^2>N$ (respectively $\frac{1}{K}\sum_{\ell=1}^K\Vert\boldsymbol{\xi}^\ell\Vert^2<N$) then $\chi_1>K$ (respectively $\chi_1<K$) and $T_c(\gamma,1)$ decreases (respectively increases) with $\gamma<\chi_1$.

Using Eqs. \eqref{eq:Ltrain}, \eqref{eq:Lgen}, for $d=1$ in the blue and green phases, gives\begin{align}
    \mathcal{L}_\text{train}-\mathcal{L}_{\text{gen}}=\begin{cases}
        \frac{1}{2}\left(\frac{T_c}{K}-\frac{T}{T_c}\right), & T<T_c(1,\gamma), \;\gamma < \chi_1,\\
        \frac{1}{2}\left(1-\frac{T}{K}\right),               &T_c(1,\gamma)<T<\chi_1^2/\gamma.
    \end{cases}
\end{align}
For the case $\frac1K\sum_{\ell=1}^K\Vert\boldsymbol{\xi}^\ell\Vert^2=N$ considered in the Main, one has $\nu_0(1)=\sqrt{K/\gamma}$, hence $T_c(1,\gamma)=K$, and thus \begin{align}
    {\cal L_\text{train}}={\cal L_\text{gen}} \qquad \text{for} \qquad T=K, \quad \gamma<K,
\end{align}as claimed.

\subsection{Test cross entropy}
We measure the ability of $P_T$ to generalize by its cross entropy relative to the data distribution $P_\star$,\begin{align} \label{eq:CE}
\mathrm{CE}(T)&=-\frac{1}{N}\left\langle\int\mathrm{d}\boldsymbol{\xi}\,P_\star(\boldsymbol{\xi})\,\ln PP_T(\boldsymbol{\xi}\vert\mathcal{D})\right\rangle_{\mathcal{D}\sim P_{\star}^K}, &\text{where}& &PP_T(\boldsymbol{\xi}\vert\mathcal{D})&=\int\mathrm{d}\boldsymbol{J}\,P_T(\boldsymbol{J}\vert\mathcal{D})P_{\sigma\vert J}(\boldsymbol{\xi}\vert\boldsymbol{J})
\end{align} is the posterior model average ($PP$ for posterior predictive). Notice that $\text{CE}(0^+)$ quantifies the performance of the MAP, see Sec.~H, while $\text{CE}(1)$ quantifies the performance of Bayesian model averaging \cite{Hoeting1999}.

Knowing the large $N$ limit of $Y$ is enough to know the large $N$ limit of $\mathrm{CE}(T)$. This is immediate for $T=1$, as the test point $\boldsymbol{\xi}$ is then entirely symmetric with the training points $\mathcal{D}$, and $PP_1$ is just the ratio of the marginal likelihood of $K+1$ data points to that of $K$ data points. This observation is readily extended to $T\neq1$ by using the trick that\begin{align} \label{eq:Xi-ratio-T}
    PP_{T}(\boldsymbol{\xi}\vert\mathcal{D})&=\frac{Y(T;\mathcal{D}';n')}{Y(T;\mathcal{D};n)}, &\text{with}& &\mathcal{D}'&=\mathcal{D}\cup\{\sqrt{T}\boldsymbol{\xi}\}, &n'&=n-1,
\end{align} since $E(\boldsymbol{\sigma};\boldsymbol{J})=T^{-1}E(T^{1/2}\boldsymbol{\sigma};\boldsymbol{J})$. In the large $N$ limit, one has \begin{align}
    \frac{1}{N}\ln PP_T(\boldsymbol{\xi}\vert\mathcal{D})=\Phi(T;n,\{\chi_k\}_{k=1}^K)-\Phi(T;n',\{\chi_k'(T)\}_{k=1}^{K+1}),
\end{align}where we have introduced the eigenvalues $\chi_k'(T)$ of the overlap matrix of $\mathcal{D}'$, defined by \begin{align}
    X_{k,l}'=\frac{1}{N}\sum_{i=1}^N \xi^{k}_i \xi^{l}_i, \quad 1\le k,l\le K, \qquad X_{K+1,K}'=X_{k,K+1}'=\frac{1}{N}\sum_{i=1}^N\xi_{i}^k\sqrt{T}\xi_{i}^{K+1}, \qquad X_{K+1,K+1}'=T.
\end{align}We see that $\sum_{k\le K+1}\chi_k'(T)=K+T$.

\section{F. Langevin sampling of the model ensembles}
\subsection{Metropolis-adjusted Langevin sampling of ensembles of spherical models}
Proposals are made according to the Euler-Maruyama discretization of the overdamped Langevin dynamics
\begin{align}
    \boldsymbol{J}'=\boldsymbol{J}_t+\tfrac{1}{2}\Delta t\,(\boldsymbol{C}-K\langle\boldsymbol{\sigma}\boldsymbol{\sigma}^\top\vert\boldsymbol{J}_t\rangle-N\gamma\boldsymbol{J})+\sqrt{2T\Delta t}\,\boldsymbol{W}_t,
    \label{eq:langevin}
\end{align}
where $\boldsymbol{W}_t$ is a symmetric matrix of identically-distributed standardized Gaussian random variables, independent up to the symmetry constraint, and $\Delta t$ is the step size. To speed up the calculation of the spin-spin correlations, we resort to the large $N$ expression $\langle \boldsymbol{\sigma} \boldsymbol{\sigma}^\top\vert\boldsymbol{J}_t\rangle \simeq(\mu_t \boldsymbol{I}-\boldsymbol{J}_t)^{-1}$, where the multiplier $\mu_t$ enforces the spherical constraint.

Proposals $\boldsymbol{J}'$ are then accepted, $\boldsymbol{J}_{t+1}=\boldsymbol{J}'$, with probability $\min\{1,\mathcal{A}(\boldsymbol{J},\boldsymbol{J}')\}$ where \begin{align}
    \mathcal{A}(\boldsymbol{J},\boldsymbol{J}')=\frac{P_T(\boldsymbol{J}'\vert\mathcal{D})\,Q(\boldsymbol{J}\vert\boldsymbol{J}')}{P_T(\boldsymbol{J}\vert\mathcal{D})\,Q(\boldsymbol{J}'\vert\boldsymbol{J})} 
\end{align} is the acceptance ratio, and $Q(\boldsymbol{Y}\vert\boldsymbol{X})\propto e^{-\mathrm{tr}(\boldsymbol{Y}-\boldsymbol{X}-\frac{1}{2}\Delta t\,(\boldsymbol{C}-K\langle\boldsymbol{\sigma}\boldsymbol{\sigma}^\top\vert\boldsymbol{J}_t\rangle-N\gamma\boldsymbol{J}))^2/4T\Delta t}$ the proposal distribution \cite{Besag1994, Roberts1996}. 

\subsection{Stochastic gradient Langevin sampling of ensembles of deep convolutional neural networks}

Following \cite{Wenzel2020}, we sample ensembles of residual neural networks, a deep convolutional neural network architecture with skip connections \cite{Bishop1995} that gained popularity due to the accuracy it achieved in image classification tasks \cite{He2015}.

We focus on the ResNet-20, a residual neural network with 20 layers known to achieve good performance on CIFAR-10. The trainable parameters are stacked into a vector $\boldsymbol{\theta}$ of size $273,258$. We define ensembles of models $\boldsymbol{\theta}$ as
\begin{align}
    P_T(\boldsymbol{\theta}\vert\{(\boldsymbol{\xi}_k,y_k)\}_{k=1}^K)\propto e^{-\frac{1}{T}U(\boldsymbol{\theta};\{(\boldsymbol{\xi}_k,y_k)\}_{k=1}^K)} \qquad \text{with} \qquad e^{-U(\boldsymbol{\theta};\{(\boldsymbol{\xi}_k,y_k)\}_{k=1}^K)}=P_G(\boldsymbol{\theta})\prod_{k=1}^K\frac{\exp(f_{\boldsymbol{\theta}}(\boldsymbol{\xi}_k)_{y_k})}{\sum_{c=1}^{10}\exp(f_{\boldsymbol{\theta}}(\boldsymbol{\xi}_k)_{c})},
\end{align}
where $y_k\in\{0,1,\dots,9\}$ labels the class of image $\boldsymbol{\xi}_k$,  $f_{\boldsymbol{\theta}}(\boldsymbol{\xi}_k)_c$ is the logit output by the network for class $c$ given image $\boldsymbol{\xi}_k$, $P_G(\boldsymbol{\theta})=\mathcal{N}(\boldsymbol{0},\gamma^{-1}\boldsymbol{I})$ with $\gamma=0.01$, and there are $K=50,000$ training examples in CIFAR-10.

To sample from $P_T(\boldsymbol{\theta}\vert\{(\boldsymbol{\xi}_k,y_k)\}_{k=1}^K)$, we use Langevin dynamics based on the unbiased estimator of the force \begin{align}
    \widehat{\boldsymbol{F}}(\boldsymbol{\theta})\equiv \frac{K}{\vert\mathcal{B}\vert}\sum_{(\boldsymbol{\xi},y)\in\mathcal{B}}\nabla_{\boldsymbol{\theta}}\log \frac{\exp(f_{\boldsymbol{\theta}}(\boldsymbol{\xi})_{y})}{\sum_{c=1}^{10}\exp(f_{\boldsymbol{\theta}}(\boldsymbol{\xi})_{c})}-\gamma\,\boldsymbol{\theta},
\end{align} where $\mathcal{B}$ is a minibatch of size $\vert\mathcal{B}\vert=128$. We simulate underdamped Langevin dynamics. The momentum obeys\begin{align}
    \boldsymbol{m}_{t+1}=(1-\zeta\Delta t)\boldsymbol{m}_t+\Delta t\,\widehat{\boldsymbol{F}}(\boldsymbol{\theta}_t)+\sqrt{2T\zeta\Delta t}\,\boldsymbol{M}^{1/2}\boldsymbol{\eta}_t,
\end{align}with friction $\zeta=14.14$, $\boldsymbol{\eta}_t\sim\mathcal{N}(\boldsymbol{0},\boldsymbol{I})$ and a mass matrix $\boldsymbol{M}$, constructed as follows. We write $\boldsymbol{\theta}=(\boldsymbol{\theta}_1,\dots,\boldsymbol{\theta}_S)$ where each $\boldsymbol{\theta}_s$ collects the $d_s$ parameters belonging to the same tensor, e.g., the convolutional kernel of layer $1$. Once every epoch, i.e., every $K/\vert\mathcal{B}\vert\approx 391$ iterations, we compute the mean squared magnitude of the rescaled force in block $s$ over $n_{\mathcal{B}}=32$ minibatches through $f_s^2=\frac{1}{n_{\mathcal{B}}d_s}\sum_{i=1}^{d_s}\sum_{j=1}^{n_{\mathcal{B}}}(\tfrac{(\widehat{F}_{s})_i^{(j)}(\boldsymbol{\theta})}{K})^2$, set $\boldsymbol{M}_s=\frac{f_s}{\min_sf_s}\boldsymbol{I}_{d_s}$, and take $\boldsymbol{M}=\text{diag}(\boldsymbol{M}_1,\dots,\boldsymbol{M}_S)$. 

The parameters are then updated through \begin{align}
    \boldsymbol{\theta}_{t+1}=\boldsymbol{\theta}_t+\Delta t\;\boldsymbol{M}^{-1}\boldsymbol{m}_{t+1},
\end{align} where $\Delta t \sim 1/\sqrt{K}$ so that updates $\boldsymbol{m}$ and $\boldsymbol{\theta}$ of order one with respect to $K$. The learning rate at epoch $\tau$ is $K(\Delta t)^2=0.1C(\tau)$. We work with cycles of $50$ epochs, in which $C(\tau)$ decays smoothly from $1$ to $0$ like a half-cosine. 

We begin sampling  a model every $50$ epochs once $150$ epochs have passed, and thus sample $n_{\boldsymbol{\theta}}=27$ models. From there, we estimate the test cross entropy loss achieved by the ensemble $P_T$ on a test set of size $P$ through \begin{align}
    \text{CE}(T)=-\frac{1}{P}\sum_{i=1}^{P}\log\frac{1}{n_{\boldsymbol{\theta}}}\sum_{j=1}^{n_{\boldsymbol{\theta}}}\frac{\exp(f_{\boldsymbol{\theta}_j}(\boldsymbol{\xi}_i)_{y_i})}{\sum_{c=1}^{10}\exp(f_{\boldsymbol{\theta}_j}(\boldsymbol{\xi}_i)_{c})}.
\end{align}
We select the $P=1,000$ test images most dissimilar to the training set. For each test image $\boldsymbol{\xi}_0$, we compute its cosine similarities to all $K=50,000$ training images in the penultimate-layer feature space of an independent ResNet-18 pretrained on ImageNet, $s_j(\boldsymbol{\xi}_0)=\boldsymbol{\phi}(\boldsymbol{\xi}_0)^\top\boldsymbol{\phi}(\boldsymbol{\xi}_k)$ $(k=1,\dots,50,000)$, where $\boldsymbol{\phi}(\boldsymbol{\xi})$ is the feature of image $\boldsymbol{\xi}$. Next, we assign to $\boldsymbol{\xi}_0$ a score $\mathcal{S}(\boldsymbol{\xi}_0)$ defined as the $90$th percentile of the $s_j(\boldsymbol{\xi}_0)$'s. We then extract the $10$th percentile $\mathcal{S}_{0.1}$ of these scores induced by all $10,000$ CIFAR-10 test images, and define the $10$\% most dissimilar images through $\mathcal{S}(\boldsymbol{\xi})\leq\mathcal{S}_{0.1}$. A test image is thus deemed dissimilar if even its relatively good matches are still not very similar.

\section{G. Low-dimensional measure of the generated data}
Suppose that a model $\boldsymbol{J}$ was sampled, such that its two largest eigenvalues $\lambda_1,\lambda_2,$ are nearly degenerate. More precisely, we assume that, in the large $N$ limit, $\lambda_{1}=\lambda+\frac{s}{2N}$, $\lambda_{2}=\lambda-\frac{s}{2N}$, with $s=N(\lambda_1-\lambda_2)=\mathcal{O}(1)$. By rotational invariance $P_{\sigma\vert J}(\boldsymbol{\sigma}\vert\boldsymbol{J})=P_{\sigma\vert J}(\widetilde{\boldsymbol{\sigma}}\vert\boldsymbol{\Lambda})$, where $\widetilde{\sigma}_{j}=\boldsymbol{\sigma}^\top \boldsymbol{v}_{j}$. We isolate the contributions of modes $i=1,2,$
$$P_{\boldsymbol{\sigma}\vert J}(\tilde{\boldsymbol{\sigma}}\vert \boldsymbol{J})\propto\delta\Big(\tilde{\sigma}_{1}^2+\tilde{\sigma}_{2}^2-N\Big(1-\tfrac{1}{N}\sum_{2<i\le N}\tilde{\sigma}_{i}^2\Big)\Big)\exp\Big(\tfrac{1}{2}\Big(\lambda_{1}\tilde{\sigma}_{1}^2+\lambda_{2}\tilde{\sigma}_{2}^2 \Big)+\tfrac{1}{2}\sum_{2<i\le N}\lambda_{i}\tilde{\sigma}_{i}^2 \Big).$$
Next, we assume that $\frac{1}{N}\sum_{2<i\le N}\tilde{\sigma}_{i}^2$ concentrates, and write $1-r^2$ its large $N$ limit. Using $\tilde{\sigma}_{1}^2+\tilde{\sigma}_{2}^2=Nr^2$,
$$\exp\left(\frac{1}{2}\left(\lambda_{1}\tilde{\sigma}_{1}^2+\lambda_{2}\tilde{\sigma}_{2}^2\right)\right)=\exp\Big(\frac{N\lambda r^2}{2}\Big)\exp\Big(\frac{s}{4N}\Big(\tilde{\sigma}_1^2-\tilde{\sigma}_2^2\Big)\Big).$$ %
Absorbing constant prefactors into the normalization, $$\rho_{2}(\tilde{\sigma}_{1},\tilde{\sigma}_{2})\propto\delta(\tilde{\sigma}_1^2+\tilde{\sigma}_2^2-Nr^2)\exp\left( \frac{s}{4N}(\tilde{\sigma}_1^2-\tilde{\sigma}_2^2)\right)$$ is the joint density of $\tilde{\sigma}_1,\tilde{\sigma}_2$, whose normalization, $z_2(r,s)$, can be computed. The result is $z(r,s)=\pi I_0(\frac{sr^2}{4})$, where $I_0(a)=\frac{1}{2\pi}\int_{0}^{2\pi}\mathrm{d}\phi\,e^{a\cos \phi}$ is the modified Bessel function of the first kind of order $0$. Switching to polar coordinates, $\tilde{\sigma}_1=\sqrt{N}r\cos\theta,\tilde{\sigma}_2=\sqrt{N}r\sin\theta$, we then have
\begin{align}
    \rho_\theta(\theta)=\frac{\exp\big(\frac{sr^2}{4}\cos(2\theta)\big)}{2\pi I_{0}\left( \frac{s r^2}{4} \right)}.
\end{align}
The uniform circular law is recovered as $s\to0$. 

Now define $\delta=(\tilde{\sigma}_1^2-\tilde{\sigma}_2^2)/N=r^2\cos(\phi)\in[-r^2,r^2]$ with $\phi=2\theta\in[0,2\pi)$. To get the density of $\delta$, note $\cos\phi = \delta /r^2$ has two solutions in $[0,2\pi)$, $\phi_0=\arccos\!\left(\delta /r^2\right)$ and $2\pi - \phi_0$. We can use $\rho_{\delta }(\delta )=\sum_{\phi:\,r^2\cos\phi=\delta }\rho_\phi(\phi)/\vert\frac{\mathrm{d}\delta }{\mathrm{d}\phi}\vert,$ where $\frac{\mathrm{d}\delta }{\mathrm{d}\phi}=-r^2\sin\phi,$ $\vert\sin\phi_0\vert=\sqrt{1-\cos^2\phi_0}=\sqrt{1-(\delta/r^2)^2}$. Summing over both solutions gives
\begin{align}
\!\rho_{\delta }(\delta )&=\frac{1}{\pi\,I_0\!\left(\frac{r^2s}{4}\right)}\,
  \frac{\exp\left(\frac{s}{4}\delta \right)}{r^2\sqrt{1-(\delta /r^2)^2}}, && \delta\in(-r^2,r^2),
\end{align}for the density of $\delta q$. The arcsine law is recovered for $s\to0$.

\begin{figure}[h]
    \centering
    \includegraphics[width=0.5\linewidth]{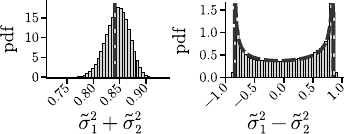}
    \caption{Learning from $K=80\sim N=100$ bump-like data of effective dimension $D=2$ with embedding dimension $N=100$ ($T=2$). Intensive mean squared projections $\widetilde{\sigma}_1^2,\widetilde{\sigma}_2^2$ of generated data onto the nearly degenerate largest pair of eigenmodes of $\boldsymbol{J}$: histograms of their \textit{(left)} sum and \textit{(left)} difference. Fluctuations around the sum are explained by finite-size effects.
    }
    \label{fig:cann}
\end{figure}

\section{H. Maximum a posteriori without replicas}
Here, we give a direct analysis of the maximum a posteriori (MAP) interaction matrix defined by \begin{align}\label{eq:MAP}
    \boldsymbol{J}_{\text{MAP}}=\underset{\boldsymbol{J}}{\arg\max}\;P_G(\boldsymbol{J})\prod_{k=1}^K\frac{e^{-E(\boldsymbol{\xi}^k;\boldsymbol{J})}}{Z(\boldsymbol{J})},
\end{align} around which $P_T(\boldsymbol{J}\vert\mathcal{D})$ concentrates in the $T\to0$ limit. We focus again on $E(\boldsymbol{\sigma};\boldsymbol{J})=-\frac{1}{2}\boldsymbol{\sigma}^\top\boldsymbol{J}\boldsymbol{\sigma}$, $\Vert\boldsymbol{\sigma}\Vert^2=N$, and $P_G(\boldsymbol{J})\propto e^{-\frac{N\gamma}{4}\mathrm{tr}(\boldsymbol{J^2})}$, $\boldsymbol{J}=\boldsymbol{J}^\top$. We write $C_{ij}=\frac{1}{K}\sum_{k=1}\xi_i^k\xi_j^k$. Eq. \eqref{eq:MAP} has a unique solution, which solves\begin{align}\label{eq:MAP-2}
    \gamma\,\boldsymbol{J}_{\mathrm{MAP}}-\tfrac{K}{N}(\boldsymbol{C}-\langle \boldsymbol{\sigma}\boldsymbol{\sigma}^\top\vert\boldsymbol{J}_{\mathrm{MAP}}\rangle)=0,
\end{align}
where $\langle \boldsymbol{\sigma}\boldsymbol{\sigma}^\top\vert\boldsymbol{J}_{\mathrm{MAP}}\rangle$ is the covariance of the data generated by $\boldsymbol{J}_{\text{MAP}}$. Let $\boldsymbol{U}$ be the matrix of eigenvectors of $\boldsymbol{C}$ with eigenvalues $c_1>\dots>c_N$, whose nonzero eigenvalues we take, for convenience, to be nondegenerate. Eq. \eqref{eq:MAP-2} says that $\boldsymbol{U}\boldsymbol{J}_{\mathrm{MAP}}\boldsymbol{U}^\top=\boldsymbol{\Lambda}_{\mathrm{MAP}}$ ($=\boldsymbol{\Lambda}$ to lighten notation) is diagonal. We introduce\begin{equation}
    q_i^2=\frac{1}{N}\left\langle\left(\boldsymbol{u}_i\cdot\boldsymbol{\sigma}\right)^2\vert\boldsymbol{J}_{\mathrm{MAP}}\right\rangle.
\end{equation} Using a Lagrange multiplier $\mu$ to impose the spherical constraint $\Vert\boldsymbol{\sigma}\Vert^2=N$, the value of $\mu$ is the unique solution in $>\lambda_{1}$ to the equation $\sum_{i=1}^Nq_i^2=\sum_{i=1}^N\frac{1}{N}\frac{1}{\mu-\lambda_{i}}=1$. In the eigenbasis of $\boldsymbol{J}_{\text{MAP}}$ and $\boldsymbol{C}$, Eq. \eqref{eq:MAP-2} becomes\begin{align}
    \label{eq:MAP-3}
    \gamma\lambda_{i}-\chi_i+Kq_i^2=0,
    \end{align}
where $q^2=\sum_{\ell=1}^dq_\ell^2=1-\sum_{d<i\le N}q_i^2$, and $\chi_j=Kc_j/N$. When $\mu-\lambda_{1}\sim\frac{1}{N}$ we allow $\lambda_{1}$ to have a near $d$-fold degeneracy, $d\le K$, in the sense that $\lambda_{\ell}-\lambda_{\ell+1}\sim\frac{1}{N}$ for $1\le\ell<d$, with $d$ to be determined.

\subsection{Undersampled regime}
In the regime \begin{align}
    N\to \infty \qquad \text{with} \qquad K =\mathcal{O}(1),
\end{align}
only $K$ out of the $N$ eigenvalues of $\boldsymbol{C}$ are nonzero, and one has $\lambda_{i}=-\frac{K}{N\gamma\mu}\to0^-$ for $K<i\leq N$, so $q^2=1-\frac{1}{\mu}$, and\begin{align}
\lambda_{\ell}&=\mu, & 1\le\ell\le d,\\
\lambda_{k}&=\frac{\chi_{k}}{\gamma}, & d<k\leq K.
\end{align}Summing Eq. \eqref{eq:MAP-3} over $\ell\le d$ to get $q^2=\sum_{\ell=1}^d\frac{\chi_\ell}{K}\left(1-\frac{\gamma\mu}{\chi_\ell}\right)$, and equating the two expressions for $q^2$, gives \begin{align}\label{eq:stieltjes}
    K\mu^{-2}-\big(K-\textstyle\sum_{\ell=1}^d\chi_\ell\big)\mu^{-1}-d\gamma&=0, &&\iff
     &\mu_{\pm}^{-1}&=\frac{1}{2K}\Big(\big(K-\textstyle\sum_{\ell=1}^d\chi_\ell\big)\pm\sqrt{\big(K-\textstyle\sum_{\ell=1}^d\chi_\ell\big)^2+4dK\gamma}\Big).
\end{align}The correct branch is $\mu=\mu_+$, since $1/\mu=\sum_{i>K}q_i^2>0$, and $\sum_{\ell=1}^d\chi_\ell\le K$ given that $\sum_{k=1}^K\chi_k=\frac{1}{N}\sum_{k=1}^K\Vert\boldsymbol{\xi}^k\Vert^2=K$. Then $K-\sum_{\ell=1}^d\chi_\ell\ge 0$ and thus $\mu_{-}^{-1}\le 0$ with equality when $dK\gamma=0$.

We can now determine the value of $d$. To start, $\mu\ge 1$ because $q^2=1-1/\mu\ge 0$. Next, since $Kq_\ell^2=\chi_\ell-\gamma\mu$, one has $q_\ell^2>0 \iff \mu<\chi_d/\gamma$ ($\ell\le d$). Finally, $\mu=\lambda_{1}=\dots=\lambda_{d}>\lambda_{d+1}=\chi_{d+1}/\gamma$ to leading order. Collecting, \begin{align}\label{eq:MAP-degeneracy}
    \max\left\{1,\frac{\chi_{d+1}}{\gamma}\right\}<\mu(d)<\frac{\chi_d}{\gamma}
\end{align}to be solved over $1\le d\le K$, with $d=0$, $\mu(0)=1$, when $\gamma>\chi_1$.

For any given $\gamma$, there is a unique $d$ satisfying Ineq. \eqref{eq:MAP-degeneracy}. To start, there is a unique $\gamma$ such that $\mu(d,\gamma)=\chi/\gamma$. Solving for $\chi=\chi_d,\chi_{d+1}$ gives that $d$ satisfies Ineq. \eqref{eq:MAP-degeneracy} iff $\gamma$ satisfies \begin{align}
    \gamma_d^+<\gamma<\gamma_d^-\nonumber,  \qquad \text{where}\qquad \gamma_d^+=\frac{\chi_{d+1}}{K}\left(K-\sum_{\ell\le d}\chi_\ell+d\chi_{d+1}\right), \qquad \gamma_d^-=\frac{\chi_{d}}{K}\left(K-\sum_{\ell\le d}\chi_\ell+d\chi_{d}\right), \quad 0\le d\le K.
\end{align} Note that $\gamma_K^+=0$ and $\gamma_1^-=\chi_1$. Using $\sum_{\ell\le d+1}\chi_\ell=\sum_{\ell\le d}\chi_\ell+\chi_{d+1}$, we obtain the identity\begin{align}
    \gamma_d^+=\gamma_{d+1}^-, \qquad d=1,\dots,K-1 \nonumber
\end{align}
for the endpoints $\gamma_{d\to d+1}$ of the $\gamma$ intervals, at which $d$ jumps by $1$, with $\gamma_{0\to1}=\chi_1$ and $\gamma_{K-1\to K}=\chi_K^2$. The line $(0,\infty)$ is thus covered by disjoint open intervals that coincide pairwise at their endpoints.

\subsection{Proportional regimes}
We consider the proportional regime \begin{align}
    N\to \infty \qquad \text{with} \qquad K=\alpha N, \qquad \alpha=\mathcal{O}(1).
\end{align}
We consider the situation where $\boldsymbol{C}$ has up to $D=\mathcal{O}(1)$ outliers, $c_m=\chi_m/\alpha$ ($m\le D$), with $N-D$ bulk eigenvalues $c_j$ ($D<j\le N$). Depending on how various quantities scale with $N$, different regimes arise, two of which we characterize.

\paragraph{$\mathcal{O}(N)+\mathcal{O}(1)$ non-extensive modes}
We begin with the regime
\begin{align}
    c_i=\mathcal{O}(1), \qquad i=1,\dots,N.
\end{align}Balance of Eq. \eqref{eq:MAP-3} imposes that $\gamma$ is at most $\mathcal{O}(1)$. Defining $a_\mu^2(c_i)=Nq_i^2(\mu)$, Eq. \eqref{eq:MAP-3} gives \begin{align}
    a_\mu^2(c_i)=\frac{1}{2\alpha}\left(\alpha c_i-\gamma\mu+\sqrt{(\alpha c_i-\gamma\mu)^2+4\alpha\gamma}\right).
\end{align} Hence all of the $q_i^2$'s are $\mathcal{O}(N^{-1})$, and there is no condensation. Equivalently\begin{align}
    \lambda_\mu(c_i)=\frac{1}{2\gamma}\left(\gamma\mu+\alpha c_i-\sqrt{(\gamma\mu-\alpha c_i)^2+4\alpha\gamma}\right).
\end{align} Thus the limiting spectral density $\rho_J$ of $\boldsymbol{J}_{\text{MAP}}$ is obtained as the image of $\rho_C$ under $c\mapsto\lambda(c)$, that is\begin{align}
    \rho_J(\lambda)&=\rho_C(c(\lambda))\left\vert\frac{\mathrm{d}c}{\mathrm{d}\lambda}\right\vert, &\text{where}& &c(\lambda)&=\frac{\gamma\lambda}{\alpha}+\frac{1}{\mu-\lambda}. 
\end{align}
Explicitly, \begin{align}
    \rho_J(\lambda)&=\begin{cases}
        \rho_J^{\text{bulk}}(\lambda), & \alpha \ge 1,\\
        (1-\alpha)\delta(\lambda-\lambda_0)+\rho_J^{\text{bulk}}(\lambda),& \alpha<1,
    \end{cases}
\end{align}with $\lambda_0=\lambda_\mu(0)$ and $\lambda_\pm=\lambda_\mu(c_\pm)$, and\begin{align}
    \rho_J^{\text{bulk}}(\lambda)=\left(\frac{\gamma}{\alpha}+\frac{1}{(\mu-\lambda)^2}\right)\rho_C^{\text{bulk}}\left(\frac{\gamma}{\alpha}\lambda+\frac{1}{\mu-\lambda}\right)\boldsymbol{1}_{\lambda\in[\lambda_-,\lambda_+]}.
\end{align}The equation fulfilled by $\mu$ can be written as\begin{align}
    1=\int\mathrm{d}c\,\rho_C(c)\,a_\mu^2(c)=\begin{cases}
        \int_{c_-}^{c_+}\mathrm{d}c\, \rho_C^{\text{bulk}}(c) \, a^2_\mu(c), &\alpha \ge 1\\
        (1-\alpha)a_\mu^2(0)+\int_{c_-}^{c_+}\mathrm{d}c\, \rho_C^{\text{bulk}}(c) \, a_\mu^2(c), &\alpha<1.
    \end{cases}
\end{align}

\begin{figure}[H]
   \centering
    \includegraphics[width=0.8\linewidth]{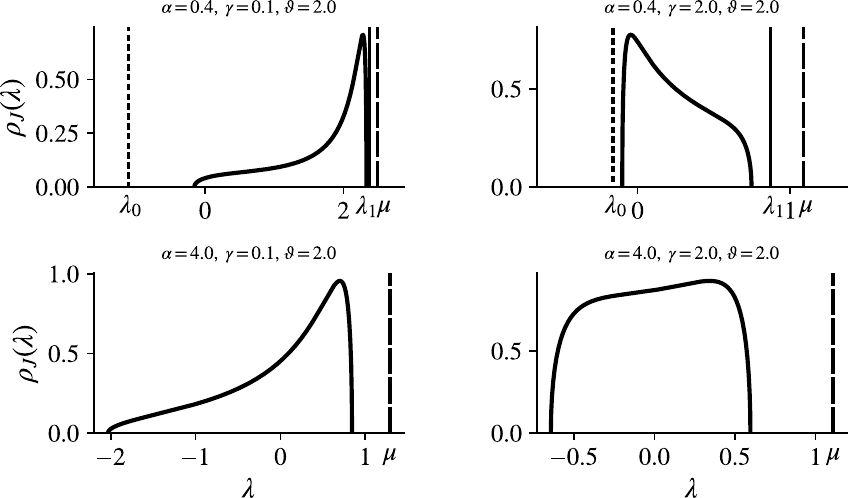}
    \caption{Eigenvalue density of $\boldsymbol{J}_{\text{MAP}}$ for spiked Wishart data ($D=1$) in the proportional regime with subextensive spikes. Although there are spectral phase transitions with eigenvalues popping out of the bulk, there is no condensation phase transition.}
    \label{fig:placeholder}
\end{figure}

\paragraph{$\mathcal{O}(N)$ non-extensive modes $+$ $\mathcal{O}(1)$ extensive modes}
We now consider the regime
\begin{align}
    c_m=\widetilde{c}_mN, \qquad \widetilde{c}_m=\mathcal{O}(1), \qquad m=1,\dots,D.
\end{align}For those modes, Eq. \eqref{eq:MAP-3} can be rewritten \begin{align}
    \gamma\lambda_m-\alpha N\widetilde{c}_m+\alpha Nq_m^2=0.
\end{align}
with $\gamma=\mathcal{O}(1), \lambda_m=\mathcal{O}(1).$ Balancing of $\mathcal{O}(N)$ terms forces \begin{align}
    \mu - \lambda_m = \mathcal{O}(N^{-1}), \qquad m=1,\dots,D.
\end{align}We then have 
\begin{align}
     q_m^2&=\widetilde{c}_m, \qquad m=1,\dots,D.
\end{align}The remaining modes $c_i=\mathcal{O}(1)$ $(i>D)$ have $q^2(c)=\mathcal{O}(N^{-1})$. 

The bulk spectral measure of $\boldsymbol{J}_{\text{MAP}}$ stays the pushforward of $\rho_C$ under the map $c\mapsto\lambda_\mu(c)$. Outliers are found at \begin{align}
    \lambda_m=\mu-\frac{1}{N\widetilde{c}_m}, \qquad m=1,\dots,D,
\end{align}where $\mu=\mathcal{O}(1)$ fulfills the implicit equation\begin{align}
    1=\sum_{m=1}^D\tilde{c}_m+\int\mathrm{d}c\;\rho_C^{\text{bulk}}(c)\;a_\mu^2(c), \qquad \text{where} \qquad \sum_{m=1}^D\widetilde{c}_m<1
\end{align}is required for the consistency of the present maximizer.

\section{I. Large $K$ asymptotics of the replica expressions}
We now come back to the replica theory, in the regime $K\to\infty$ assuming that \begin{align}
    \chi_k=\begin{cases}
        \widetilde{\chi}_kK, & 1\le k\le D,\\
        \sigma_Kx_k, &D<k\le K,
    \end{cases} \qquad \widetilde{\chi}_k,x_k=\mathcal{O}(1), \qquad \sigma_K = o(K).
\end{align} 
Extensive modes always lead to condensation if $\chi_\ell/\gamma\to\infty$. No spurious direction is recovered for $\gamma$ large enough, \begin{align}
   \gamma>\gamma_c \qquad \text{with}\qquad \gamma_c=\chi_{D+1}\delta_K+\frac{D}{K}\chi_{D+1}^2.
\end{align}
Inspection of the expression for $\nu_0(d)$ reveals that different regimes arise depending on the amplitude of \begin{align}
    \delta_K=1-\sum_{k\le D}\widetilde{\chi}_k
\end{align}
which must be compared to $\sqrt{\gamma/K}$:
\vskip .3cm
\noindent \textit{(i.a) Regime} $\delta_K\ll \sqrt{\gamma/K}$. 
\begin{align}
     \mu&=\frac{K}{D\gamma}-\frac{K\delta_K}{2D\gamma}+o\left(\frac{K\vert\delta_K\vert}{\gamma}\right),\nonumber\\
     \nu_0(D)&=\sqrt{\frac{D\gamma}{K}}+\frac{\delta_K}{2}+\mathcal{O}\left(\frac{\delta_K^2}{\sqrt{\gamma/K}}\right),\nonumber\\
     \frac{m_\ell^2}{n}&=\widetilde{\chi}_\ell-\sqrt{\frac{\gamma}{DK}}+\frac{\delta_K}{2D}+o\left(K^{-1/2}\right),\\
     T_c&=\frac{K}{D}-\frac{K^{3/2}\delta_K}{D^{3/2}\sqrt{\gamma}}+o(K^{3/2}\delta_K).
\end{align}
\vskip .3cm
\noindent \textit{(i.b) Regime} $\delta_K\sim a\sqrt{\gamma/K}$. 
\begin{align}
     \mu&=\sqrt{\frac{K}{\gamma}}\frac{2}{a+\sqrt{a^2+4D}}+\frac{T}{2\sqrt{K\gamma}}(a+\sqrt{a^2+4D})+o(K^{-1/2}),\nonumber\\
     \nu_0(D) &= \sqrt{\frac{\gamma}{K}}\frac{1}{2}\left(a+\sqrt{a^2+4D}\right)+o(K^{-1/2}),\nonumber\\ 
     \frac{m_\ell^2}{n}&=\widetilde{\chi}_\ell-\sqrt{\frac{\gamma}{K}}\frac{2}{a^2+\sqrt{a^2+4D}}+o(K^{-1/2}),\nonumber\\
     T_c&=\frac{4K}{(a+\sqrt{a^2+4D})^2}+o(K).
\end{align}

\vskip .3cm
\noindent \textit{(ii) Regime} $\delta_K\gg \sqrt{\gamma/K}$. 
\begin{align}
    \mu&=\frac{1}{\delta_K}+\frac{T}{\gamma}\delta_K+\frac{D}{K}\left(\frac{T}{\delta_K}-\frac{\gamma}{\delta_K^2}\right)+\mathcal{O}\left(\frac{\gamma^2}{K^2\delta_K^5}\right),\nonumber\\
    \nu_0(D) &=\delta_K+\frac{D\gamma}{K\delta_K}+\mathcal{O}\left(\frac{\gamma^2}{K^2\delta_K^3}\right), \nonumber\\
    \frac{m_\ell^2}{n}&=\widetilde{\chi}_\ell-\frac{\gamma}{K\delta_K}+\mathcal{O}\left(\frac{\delta_K}{K}\right), \nonumber\\
    T_c&=\frac{\gamma}{\delta_K^2}-\frac{2D\gamma^2}{K\delta_K^4}+\mathcal{O}\left(\frac{\gamma^3}{K^2\delta_K^6}\right).
\end{align}
Here the prediction for $\mu$ is the same as in MAP only for $\delta_K\ll T/\gamma$.

In the large $K$ limit, the leading $\mathcal{O}(K^2)$ contribution to the saddle point value of $\Phi$, Eq. \eqref{eq:log-marginal-likelihood}, vanishes.

\section{J. Replica calculation in the proportional regime}
We consider $Y=\overline{Z(\boldsymbol{J})^n}$ again, only this time in the regime \begin{align}
    \chi_k=\begin{cases}
        K\widetilde{\chi}_k, & 1\le k\le D \\
        \mathcal{O}(1), & D<k\le K,
    \end{cases} \qquad N\to\infty, \qquad K=\alpha N, \qquad \qquad D, \alpha, \widetilde{\chi}_k=\mathcal{O}(1).
\end{align}
We argue that the integral $\overline{Z(\boldsymbol{J})^n}$ is dominated by matrices that are simple, i.e., characterized by an $\mathcal{O}(1)$ number of $O(n)$ invariants whose amplitudes are $\mathcal{O}(N)$. We show that, as a result, this integral is dominated by an $\mathcal{O}(N^3)$ energetic contribution, while entropic terms and fluctuations around the saddle point remain $\mathcal{O}(N^2)$.

We start from $\overline{Z(\boldsymbol{J})^n}\propto\int\mathrm{d}(\boldsymbol{Q},\widehat{\boldsymbol{Q}},\boldsymbol{M},\widehat{\boldsymbol{M}})\,\exp(N \Phi(\boldsymbol{Q},\widehat{\boldsymbol{Q}},\boldsymbol{M},\widehat{\boldsymbol{M}}))$ where, up to irrelevant additive constants,
\begin{align}
\Phi&=\frac{T}{4\gamma}\mathrm{tr}(\boldsymbol{Q}^2)+\frac{1}{2\gamma}\mathrm{tr}(\boldsymbol{M}\boldsymbol{D}_\chi\boldsymbol{M}^\top)-\frac{1}{2}\mathrm{tr}(\boldsymbol{Q}\widehat{\boldsymbol{Q}})-\mathrm{tr}(\boldsymbol{M}\widehat{\boldsymbol{M}}^\top)-\frac{1}{2}\mathrm{tr}(\widehat{\boldsymbol{M}}^\top\widehat{\boldsymbol{Q}}^{-1}\widehat{\boldsymbol{M}})+\frac{1}{2}\ln\det(-\widehat{\boldsymbol{Q}}^{-1}).
\end{align}
Diagonalizing $\widehat{\boldsymbol{Q}}=\widehat{\boldsymbol{\Omega}}\,\text{diag}(\widehat{\boldsymbol{\nu}})\,\widehat{\boldsymbol{\Omega}}^\top$, $\boldsymbol{Q}=\boldsymbol{\Omega}\,\text{diag}(\boldsymbol{\nu})\,\boldsymbol{\Omega}^\top$, and rotating, $\widetilde{\boldsymbol{M}}=\boldsymbol{\Omega}^{\!\top}\!\boldsymbol{M}$, $\widetilde{\widehat{\boldsymbol{M}}}=\widehat{\boldsymbol{\Omega}}^{\!\top}\!\widehat{\boldsymbol{M}}$, we obtain the expression
\begin{align}
    \overline{Z(\boldsymbol{J})^n}&\propto \int\mathrm{d}(\boldsymbol{\nu},\widehat{\boldsymbol{\nu}},\widetilde{\boldsymbol{M}},\widetilde{\widehat{\boldsymbol{M}}})\,\Delta(\boldsymbol{\nu})\,\Delta(\widehat{\boldsymbol{\nu}})\,\mathcal{J}_n(\boldsymbol{\nu},\widehat{\boldsymbol{\nu}})\,\exp(N\mathcal{G}(\boldsymbol{\nu},\widehat{\boldsymbol{\nu}},\widetilde{\boldsymbol{M}},\widetilde{\widehat{\boldsymbol{M}}})),\\
    \mathcal{J}_n(\boldsymbol{\nu},\widehat{\boldsymbol{\nu}})&=\int_{O(n)}\mathcal{D}\boldsymbol{\Omega}\;\delta\!\left(\sum_{b=1}^n\nu_b\Omega_{ab}^2-1\right)\int_{O(n)}\mathcal{D}\widehat{\boldsymbol{\Omega}}\,\exp\Big(-\frac{N}{2}\sum_{ab}\nu_a\widehat{\nu}_bR_{ab}^2-N\sum_{abk}\widetilde{M}_{ak}\widetilde{\widehat{M}}_{bk}R_{ab}\Big),\\
    \mathcal{G}(\boldsymbol{\nu},\widehat{\boldsymbol{\nu}},\widetilde{\boldsymbol{M}},\widetilde{\widehat{\boldsymbol{M}}})&=\frac{T}{4\gamma}\sum_{a}\nu_a^2+\frac{1}{2\gamma}\sum_{ak} \chi_k\widetilde{M}_{ak}^2-\frac{1}{2}\sum_{ak}\widehat{\nu}_a^{-1}\widetilde{\widehat{M}}_{ak}^2+\frac{1}{2}\sum_a\ln-\widehat{\nu}_a^{-1},
\end{align}
where $\Delta(\boldsymbol{\nu}),\Delta(\widehat{\boldsymbol{\nu}})$, are Vandermonde determinants, see Eq. \eqref{eq:nu-Vandermonde}, and $\boldsymbol{R}=\boldsymbol{\Omega}^\top\widehat{\boldsymbol{\Omega}}$ is the relative rotation between the eigenbases of $\boldsymbol{Q}$ and $\widehat{\boldsymbol{Q}}$.

We assume that the solution found by sending $N\to\infty$ while holding $K$ fixed, and then taking $K\to\infty$, still holds,
\begin{align}
    \nu_a&=\begin{cases}
        K\widetilde{\nu}_a, & 1\le a\le D,\\
        \mathcal{O}(1), & D<a\le n,
    \end{cases} &\widehat{\nu}_a&=\begin{cases}
        K\widetilde{\widehat{\nu}}_a& 1\le a\le D,\\
        \mathcal{O}(1), & D<a\le n,
    \end{cases} 
    &\widetilde{\nu}_a, \widetilde{\widehat{\nu}}_a&=\mathcal{O}(1),
    \\ 
    \widetilde{M}_{ak}&=\begin{cases}
        \delta_{ak}\,K^{1/2}\,\widetilde{m}_k, & 1\le a,k\le D,\\
        \mathcal{O}(K^{-1/2}), & D<a,k\le n,\\
    \end{cases} 
    &\widetilde{\widehat{M}}_{ak}&=\begin{cases}
        \delta_{ak}\,K^{3/2}\,\widetilde{\widehat{m}}_k, & 1\le a,k\le D,\\
        \mathcal{O}(K^{-1/2}), & D<a,k\le n,
    \end{cases} &\widetilde{m}_k, \widetilde{\widehat{m}}_k&=\mathcal{O}(1),
\end{align}
see Secs.~A and ~I. This Ansatz is consistent with numerical simulations of Monte Carlo sampling of the posterior, see Sec.~F and Fig.~5(c, d) in Main. We now determine the scaling of the various contributions to $\overline{Z(\boldsymbol{J})^n}$ under this Ansatz.

Starting with $\mathcal{G}$, we see that $\mathcal{G}=K^2\mathcal{G}^{(2)}+\mathcal{O}(K)$ 
where
\begin{align}
    \mathcal{G}^{(2)}=\frac{T}{4\gamma}\sum_{k\le D}\widetilde{\nu}_k^2+\frac{1}{2\gamma}\sum_{k\le D} \widetilde{\chi}_k\widetilde{m}_{k}^2-\frac{1}{2}\sum_{k\le D}\widetilde{\widehat{\nu}}_k^{-1}\widetilde{\widehat{m}}_{k}^2. 
\end{align} 
Hence $e^{N\Phi}\sim e^{\mathcal{O}(NK^2)}$.

For the Vandermonde determinants, we have \begin{align}
    \log\Delta(\boldsymbol{\nu})&=\sum_{1\le k<\ell\le D}\log \vert\nu_k-\nu_\ell\vert+\sum_{k\le D<a}\log\vert\nu_k-\nu_a\vert+\sum_{D<a<b\le n}\log\vert\nu_a-\nu_b\vert=\mathcal{O}(\log K)+\mathcal{O}(K\log K)+\mathcal{O}(K^2)\nonumber
\end{align}and similarly for $\log\Delta(\widetilde{\boldsymbol{\nu}})$. Thus $\Delta(\boldsymbol{\nu}),\Delta(\widetilde{\boldsymbol{\nu}})\sim e^{\mathcal{O}(K^2)}$.

We now deal with $\mathcal{J}_n$, an instance of matrix spherical integral. We make the change of variables 
$\widehat{\boldsymbol{\Omega}}\mapsto\boldsymbol{R}$. The Haar measure on $O(n)$ is invariant under multiplication by orthogonal matrices. By Haar invariance $\mathcal{J}_n$ thus factorizes as
\begin{align}
    \mathcal{J}_n(\boldsymbol{\nu},\widehat{\boldsymbol{\nu}})&=\mathcal{I}_n(\boldsymbol{\nu})\int_{O(n)}\mathcal{D}\boldsymbol{R}\,\exp\Big(-\frac{N}{2}\sum_{ab}\nu_a\widehat{\nu}_bR_{ab}^2-N\sum_{abk}\widetilde{M}_{ak}\widetilde{\widehat{M}}_{bk}R_{ab}\Big),\nonumber
\end{align}
where $\mathcal{I}_n(\boldsymbol{\nu})$ is the already introduced matrix spherical integral, Eq. \eqref{eq:I_n}, which imposes the spherical constraints.

We expect $\mathcal{I}_n\sim e^{\mathcal{O}(K^2)}$ for this purely entropic integral. The reason is that $\boldsymbol{\Omega}$ consists in $n^2$ parameters, and orthonormality $\boldsymbol{\Omega}^{\!\top}\boldsymbol{\Omega}=\boldsymbol{I}_n$ is imposed with $n(n+1)/2$ independent constraints, giving $\dim O(n)=n(n-1)/2=\mathcal{O}(K^2)$.

The nontrivial integral is $\mathcal{J}_n(\boldsymbol{\nu
},\widetilde{\boldsymbol{\nu}})/\mathcal{I}_n(\boldsymbol{\nu})$. Writing the Haar measure more explicitly \begin{align}
    \mathcal{D}\boldsymbol{R}\propto \Big[\prod_{a,b=1}^n\mathrm{d}R_{ab}\Big]\prod_{1\le a\le b\le n}\delta\!\left(\sum_{c=1}^nR_{ac}R_{bc}-\delta_{ab}\right),\nonumber
\end{align}
these constraints are imposed using Lagrange multipliers $W_{ab}=W_{ba}$,\begin{align}
    \delta\!\left(\sum_{c=1}^nR_{ac}R_{bc}-\delta_{ab}\right)\propto \int \mathrm{d}W_{ab}\;\exp\Big(NW_{ab}\delta_{ab}-NW_{ab}\sum_{c=1}^nR_{ac}R_{bc}\Big).\nonumber
\end{align} Denoting $\mathrm{d}\boldsymbol{R}=\prod_{ab}\mathrm{d}R_{ab}$, $\mathrm{d}\boldsymbol{W}=\prod_{a\le b}\mathrm{d}W_{ab}\;\boldsymbol{1}_{W_{ab}=W_{ba}}$, we thus have \begin{align}
    \mathcal{J}_n(\boldsymbol{\nu},\widehat{\boldsymbol{\nu}})&=\int\mathrm{d}\boldsymbol{R}\;\mathrm{d}\boldsymbol{W}\,e^{N\mathcal{H}(\boldsymbol{R},\boldsymbol{W};\boldsymbol{\nu},\widehat{\boldsymbol{\nu}})},\\
    \mathcal{H}(\boldsymbol{R},\boldsymbol{W};\boldsymbol{\nu},\widehat{\boldsymbol{\nu}})&=\frac{1}{2}\sum_aW_{aa}-\frac{1}{2}\sum_{abc}W_{ab}R_{ac}R_{bc}-\frac{1}{2}\sum_{ab}\nu_a\widehat{\nu}_bR_{ab}^2-\sum_{abk}\widetilde{M}_{ak}\widetilde{\widehat{M}}_{bk}R_{ab}.\nonumber
\end{align}
We begin with terms that do not depend on $W_{ab}$. For the quadratic term in $R_{ab}$, we write
\begin{align}
    \Big\vert\sum_{ab}\nu_a\widehat{\nu}_bR_{ab}^2\Big\vert\le\sum_{ab}\vert\nu_a\vert\vert\widehat{\nu}_b\vert R_{ab}^2=\Big(\sum_{a,b\le D}+\sum_{a\le D}\sum_{D<b\le n}+\sum_{D<a\le n}\sum_{b\le D}+\sum_{D<a,b\le n}\Big)\vert\nu_a\vert\vert\widehat{\nu}_b\vert R_{ab}^2,\nonumber
\end{align} and, using $\sum_bR_{ab}^2=\sum_{a}R_{ab}^2=1$, we note that
\begin{align}
    \sum_{a,b\le D}\vert\nu_a\vert\vert\widehat{\nu}_b\vert R_{ab}^2&=\mathcal{O}(K^2), &
    \sum_{a\le D}\sum_{D<b\le n}\vert\nu_a\vert\vert\widehat{\nu}_b\vert R_{ab}^2&=\mathcal{O}(K),\nonumber\\
    \sum_{D<a\le n}\sum_{b\le D}\vert\nu_a\vert\vert\widehat{\nu}_b\vert R_{ab}^2&=\mathcal{O}(K), &
    \sum_{D<a,b\le n}\vert\nu_a\vert\vert\widehat{\nu}_b\vert R_{ab}^2&=\mathcal{O}(K).\nonumber
\end{align}
Next, for the linear term in $R_{ab}$, \begin{align}
    \sum_{abk}\widetilde{M}_{ak}\widetilde{\widehat{M}}_{bk}R_{ab}=K^2\sum_{k\le D}\widetilde{m}_k\widetilde{\widehat{m}}_kR_{kk}+o(K^2).\nonumber
\end{align}
Above we wrote $\boldsymbol{R}$ as \begin{align}
    \boldsymbol{R}=\begin{pmatrix}
        \boldsymbol{R}^{11}& \boldsymbol{R}^{12}\\
        (\boldsymbol{R}^{12})^\top & \boldsymbol{R}^{22}
    \end{pmatrix} \qquad \text{with} \qquad \boldsymbol{R}^{11}\in \mathbb{R}^{D\times D}.\nonumber
\end{align}Notice that $\boldsymbol{R}\in O(n)$ implies $\boldsymbol{R}^{11}(\boldsymbol{R}^{11})^\top+\boldsymbol{R}^{12}(\boldsymbol{R}^{12})^\top=\boldsymbol{I}_D$. For the leading order saddle to meet the latter constraint, rescale the multipliers, $W_{k\ell}=K^2\widetilde{W}_{k\ell}, \qquad \widetilde{W}_{k\ell}=\mathcal{O}(1)$,  ($k,\ell \le D$). Combining everything, we have $\mathcal{H}=K^2\mathcal{H}^{(2)}+o(K^2)$ with
\begin{align}
    \mathcal{H}^{(2)}=\frac{1}{2}\sum_{k\le D}\widetilde{W}_{kk}-\frac{1}{2}\sum_{k,\ell,m\le D}\widetilde{W}_{k\ell}R_{km}R_{\ell m}-\frac{1}{2}\sum_{k,\ell}\widetilde{W}_{k\ell}G_{k\ell}-\frac{1}{2}\sum_{k,\ell\le D}\nu_k\widehat{\nu}_\ell R_{k\ell}^2-\sum_{k\le D}\widetilde{m}_k\widetilde{\widehat{m}}_kR_{kk},\nonumber
\end{align}where $\mathcal{H}^{(2)}$ depends on $R_{kc}$ ($k\le D,c>D$) through $G_{k\ell}=\sum_{c>D}R_{kc}^{}R_{\ell c}$. The associated entropic contribution is
\begin{align}
    \mathcal{N}(\boldsymbol{G})=\int\Big[\prod_{1\le k\le D<c\le n}\mathrm{d}R_{kc}\Big]\;\prod_{1\le k\le\ell\le D}\delta\!\left(\sum_{D<c\le n}R_{kc}R_{\ell c}-G_{k\ell}\right)\propto (\det\boldsymbol{G})^{(n-2D-1)/2}.\nonumber
\end{align}This follows by making the change of variables $\boldsymbol{R}'=\boldsymbol{G}^{1/2}\boldsymbol{R}$. Because $\mathrm{tr}(\boldsymbol{G})\le D=\mathcal{O}(1)$, we expect $\mathcal{N}(\boldsymbol{G})\le e^{\mathcal{O}(K)}$.

We are left with the following integral, which we may evaluate by the saddle point method,
\begin{align}
    &\overline{Z(\boldsymbol{J})^n}\propto\int \Big[\prod_{k=1}^D\mathrm{d}\widetilde{\nu}_k\,\mathrm{d}\widetilde{\widehat{\nu}}_k\,\mathrm{d}\widetilde{m}_{k}\,\mathrm{d}\widetilde{\widehat{m}}_{k}\Big]\Big[\prod_{1\le k,\ell\le D}\mathrm{d}R_{k\ell}\Big]\Big[\prod_{1\le k\le\ell\le D}\mathrm{d}G_{k\ell}\,\mathrm{d}\widetilde{W}_{k\ell}\Big]\;e^{\alpha^2N^3\mathcal{S}(\widetilde{W}_{k\ell},R_{k\ell},G_{k\ell},\widetilde{\nu}_k,\widetilde{\widehat{\nu}},\widetilde{m}_{k},\widetilde{\widehat{m}}_{k})+\mathcal{O}(N^2)},\nonumber
\end{align}
where $\mathcal{S}(\widetilde{W}_{k\ell},R_{k\ell},G_{k\ell},\widetilde{\nu}_k,\widetilde{\widehat{\nu}},\widetilde{m}_{k},\widetilde{\widehat{m}}_{k})=\mathcal{G}^{(2)}(\widetilde{\nu}_k,\widetilde{\widehat{\nu}}_k,\widetilde{m}_{k},\widetilde{\widehat{m}}_{k})+\mathcal{H}^{(2)}(\widetilde{W}_{k\ell},R_{k\ell},G_{k\ell};\nu_k,\widehat{\nu}_k,\widetilde{m}_{k},\widetilde{\widehat{m}}_{k})$.

Stationary points of $\mathcal{S}$ satisfy
\begin{align}
  0&=\begin{cases}
     \frac{\partial\mathcal{H}^{(2)}}{\partial G_{k\ell}},\\
     \frac{\partial\mathcal{H}^{(2)}}{\partial \widetilde{W}_{k\ell}},\\
     \frac{\partial\mathcal{H}^{(2)}}{\partial R_{k\ell}},
   \end{cases} \quad \iff \qquad 0 =\begin{cases}
       \widetilde{W}_{k\ell},\\
       \delta_{k\ell}-G_{k\ell}-\sum_{m=1}^DR_{km}R_{\ell m},\\
       \widetilde{\nu}_k\widetilde{\widehat{\nu}}_\ell R_{k\ell}+\delta_{k\ell}\widetilde{m}_k\widetilde{\widehat{m}}_k, 
   \end{cases} && (k,\ell \le D),\\
    0&=\begin{cases}
      \frac{\partial \mathcal{S}^{(2)}}{\partial\widetilde{\nu}_k},\\
      \frac{\partial \mathcal{S}^{(2)}}{\partial\widetilde{\widehat{\nu}}_k},\\
      \frac{\partial \mathcal{S}^{(2)}}{\partial\widetilde{m}_k},\\
      \frac{\partial \mathcal{S}^{(2)}}{\partial\widetilde{\widehat{m}}_k},
    \end{cases} \quad \iff \qquad 0=\begin{cases}
      \frac{T}{\gamma}\widetilde{\nu}_k-\sum_{\ell=1}^D\widetilde{\widehat{\nu}}_\ell R_{k\ell}^2,\\
      \widetilde{\widehat{m}}_k^2-\widetilde{\widehat{\nu}}_k^2\sum_{\ell=1}^D\widetilde{\nu}_\ell R_{\ell k}^2,\\
      \frac{\chi_k}{\gamma}\widetilde{m}_k-\widetilde{\widehat{m}}_kR_{kk},\\
      \widetilde{\widehat{m}}_k+\widetilde{\widehat{\nu}}_k\widetilde{m}_kR_{kk},
    \end{cases} && (k, \ell \le D).
\end{align}From the $R_{k\ell}$ equation, we get \begin{align}
    &R_{k\ell}=0,  && \text{for} && \widetilde{\nu}_k,\widetilde{\widehat{\nu}}_\ell\neq0, &&1\le k<\ell\le D,
\end{align}which combined with the $\widetilde{W}_{k\ell}$ equation gives \begin{align}
    G_{k\ell}&=0, &&(1\le k<\ell\le D), &&
    &&G_{kk}=1-R_{kk}^2, &&(k\le D).
\end{align}
Combining these with the conditions for the stationarity $\mathcal{S}^{(2)}$, we obtain the relations
\begin{align}
    &\widetilde{\nu}_k=-\frac{\widetilde{\chi}_k}{T}, &&\widetilde{m}_k^2=\widetilde{\nu}_k, &&\widetilde{\widehat{\nu}}_kR_{kk}^2=-\frac{\widetilde{\chi}_k}{\gamma}, && \widetilde{\widehat{m}}_kR_{kk}=\frac{\widetilde{\chi}_k}{\gamma}\widetilde{m}_k.
\end{align}
It remains to show that maximizers satisfy $G_{kk}\downarrow 0$. The leading order is independent of $G_{kk}$. We must consider the  entropy $\log \mathcal{N}(\boldsymbol{G})$. Since the matrices $\boldsymbol{G}$ making the leading order in the exponent stationary are diagonal, one has
\begin{align}
    \log \mathcal{N}(\boldsymbol{G})=\frac{n-2D-1}{2}\sum_{k=1}^D\log(1-R_{kk}^2),\nonumber
\end{align}which is positive since $n<0$ and $0<R_{kk}^2<1$. It is maximized at the boundary $G_{kk}\downarrow 0$, i.e., $R_{kk}^2\uparrow1$, hence \begin{align}
    G_{kk}=0, \qquad R_{kk}^2=1, \qquad (k\le D).
\end{align}
We thereby recover
\begin{align}
    \widetilde{\nu}_k=\widetilde{m}_k^2=-\frac{\widetilde{\chi}_k}{T}, \qquad \widetilde{\widehat{\nu}}_k=\frac{T}{\gamma}\widetilde{\nu}_k, \qquad \widetilde{\widehat{m}}_k^2=\left(\frac{\widetilde{\chi}_k}{\gamma}\right)^{\!\!2}\widetilde{m}_k^2.
\end{align}Lastly, enforcing the soft spherical constraint yields, to leading order,\begin{align}
    \nu_0\equiv\frac{1}{n-D}\sum_{D<a\le n}\nu_a=1-\sum_{k=1}^D\widetilde{\chi}_k\,.
\end{align}

\bibliography{refs}